\documentclass[12pt,number,sort&compress]{elsarticle}
\usepackage[margin=1in]{geometry}
\usepackage{caption}
\usepackage{subcaption}
\usepackage{amsmath}
\usepackage[version=3]{mhchem}
\usepackage[dvipsnames]{xcolor}
\usepackage{xr-hyper}
\usepackage[colorlinks=true,linkcolor=black, citecolor=blue, urlcolor=blue]{hyperref}
\definecolor{darkgreen}{rgb}{0.0,0.5,0.0}

\makeatletter
\newcommand*{\addFileDependency}[1]{
  \typeout{(#1)}
  \@addtofilelist{#1}
  \IfFileExists{#1}{}{\typeout{No file #1.}}
}
\makeatother

\newcommand*{\myexternaldocument}[1]{%
    \externaldocument{#1}%
    \addFileDependency{#1.tex}%
    \addFileDependency{#1.aux}%
}
\myexternaldocument{supplement}
\journal{npj Computational Materials}

\begin{document}

\title{Multi-scale Investigation of Chemical Short-Range Order and Dislocation Glide in the  MoNbTi and TaNbTi Refractory Multi-Principal Element Alloys}

\author[UCSD]{Hui Zheng \fnref{cor2}}
\author[UCSB]{Lauren T.W.\ Fey \fnref{cor2}}
\author[UCSD]{Xiang-Guo Li \fnref{cor2}}
\fntext[cor2]{These authors contributed equally}

\author[UM,Drexel]{Yong-Jie Hu}
\author[UM]{Liang Qi}
\author[UCSD]{Chi Chen}

\author[UCSBME]{Shuozhi Xu}
\author[UCSB,UCSBME]{Irene J.\ Beyerlein \corref{cor1}}
\ead{beyerlein@ucsb.edu}

\author[UCSD]{Shyue Ping Ong \corref{cor1}}
\ead{ongsp@eng.ucsd.edu}
\cortext[cor1]{Corresponding author}

\address[UCSD]{Department of NanoEngineering, University of California San Diego, La Jolla, CA 92093-0448, USA}
\address[UCSB]{Materials Department, University of California, Santa Barbara, CA 93106-5050, USA}
\address[UCSBME]{Department of Mechanical Engineering, University of California, Santa Barbara, CA 93106-5070, USA}
\address[UM]{Department of Materials Science and Engineering, University of Michigan, Ann Arbor, MI 48109, USA}
\address[Drexel]{Department of Materials Science and Engineering, Drexel University, Philadelphia, PA 19104, USA}
\date{}

\begin{abstract}
Refractory multi-principal element alloys (RMPEAs) are promising materials for high-temperature structural applications. Here, we investigate the role of chemical short-range ordering (CSRO) on dislocation glide in two model RMPEAs --- TaNbTi and MoNbTi --- using a multi-scale modeling approach. A highly accurate machine learning interatomic potential was developed for the Mo-Ta-Nb-Ti system and used to demonstrate that MoNbTi exhibits a much greater degree of SRO than TaNbTi and the local composition has a direct effect on the unstable stacking fault energies (USFE). From mesoscale phase-field dislocation dynamics simulations, we find that increasing SRO leads to higher mean USFEs, thereby increasing the stress required for dislocation glide. The gliding dislocations experience significant hardening due to pinning and depinning caused by random compositional fluctuations, with higher SRO decreasing the degree of USFE dispersion and hence, amount of hardening. Finally, we show how the morphology of an expanding dislocation loop is affected by the applied stress, with higher SRO requiring higher applied stresses to achieve smooth screw dislocation glide. 

\end{abstract}

\maketitle

\section{Introduction}

Refractory multi-principal element alloys (RMPEAs), which are composed of refractory metals in highly concentrated proportions, have garnered intense interest as promising candidate materials for high-temperature structural applications \cite{senkovMicrostructurePropertiesRefractory2015, sheikhAlloyDesignIntrinsically2016}. Most RMPEAs studied to date possess superior yield strengths and several RMPEAs, such as TiZrHfNbTa, TaNbTi, and HfNbTa, have reported good tensile ductility as well \cite{dirrasElasticPlasticProperties2016, georgeHighentropyAlloys2019a, georgeHighEntropyAlloys2020,wangNewTernaryEquiatomic2019}. To accelerate design of structural RMPEAs within the vast composition space, it is necessary to understand the dislocation-based mechanisms driving their response to applied load. RMPEAs are chemically disordered at the atomic scale leading to unusual dislocation behavior. Several molecular dynamics (MD) simulations and transmission electron microscopy studies alike have pointed to stochastic glide, slip on higher-order glide planes, screw dislocation cross-kinking, and relatively low edge dislocation mobility at higher frequencies in RMPEAs than expected in pure refractories \cite{raoModelingSolutionHardening2019b,wangMultiplicityDislocationPathways2020,chenUnusualActivatedProcesses2020a,leeStrengthCanBe2021,raoTheorySolidSolution2021,xu_line-length-dependent_2022}.


In most studies and analyses, the chemical disorder was treated as being ideally uniformly random exhibiting no thermodynamically driven chemical short-range order (CSRO). While this may be a suitable approximation at high temperatures or for the bulk, some preferential pairing between elements in the alloy, or SRO, can be expected to manifest, especially over length scales corresponding to a dislocation.  In recent years, detectable levels of SRO have been reported in both FCC MPEAs such as CoCrNi \cite{zhangLocalStructureShortRange2017a, dingTunableStackingFault2018d} and CoNiV \cite{chenDirectObservationChemical2021} as well as the BCC RMPEAs such as HfNbZr \cite{guoLocalAtomicStructure2013}. Supporting calculations of SRO in MPEAs have been provided via density functional theory (DFT) and hybrid Monte Carlo (MC)/MD simulations, where temperature is used to adjust the extent and severity of SRO \cite{singhAtomicShortrangeOrder2015,wrobelPhaseStabilityTernary2015,fernandez-caballeroShortRangeOrderHigh2017,maChemicalShortrangeOrders2018,jianEffectsLatticeDistortion2020,zhaoLocalOrderingTendency2021,jian_role_2021,xie_role_2021}. DFT calculations have shown that SRO in MoNbTaW RMPEA reduces the variation in the dislocation core properties, while have little effect on their core energies \cite{yinInitioModelingEnergy2020}. MC/MD simulations of nanocrystalline FCC and RMPEAs predict a preference for greater degrees of SRO near the grain boundaries than the interiors and overall greater strength \cite{liComplexStrengtheningMechanisms2020a, jianEffectsLatticeDistortion2020}. In MD simulations of high-velocity dislocation motion, SRO tended to increase edge dislocation mobility but decrease screw dislocation mobility in RMPEA \cite{yinAtomisticSimulationsDislocation2021}, while the partial dislocation mobilities were decreased in an FCC MPEA \cite{jianEffectsLatticeDistortion2020}.

In this work, we seek to elucidate the extent of CSRO affects the motion of dislocations, in both mechanically under-driven and over-driven situations. We introduce a multiscale modeling approach that links a new machine learning potential with quantum fidelity to hybrid MC/MD for temperature-dependent CSRO calculations at the atomic scale and ultimately to dislocation dynamics simulation of long dislocations and dislocation loops at the mesoscale. The interatomic potential presented here is a highly accurate one developed for the Mo-Ta-Nb-Ti system \cite{chenAccurateForceField2017b,liQuantumaccurateSpectralNeighbor2018b,zuoPerformanceCostAssessment2020c,qiBridgingGapSimulated2021,yinAtomisticSimulationsDislocation2021}. With it, two base ternary systems, MoNbTi and TaNbTi, are extracted and studied via hybrid MC/MD for CSRO at two annealing temperatures. We elected these MPEAs since experimental observations find that their equimolar forms exhibit disparate mechanical properties, with the MoNbTi being substantially greater in tensile yield strength, peak strength and strain hardening than TaNbTi. To study the effect of CSRO on the dislocation glide mechanisms, a method is developed to build statistical instantiations of large 3D crystals with spatial mapping of CSRO and associated unstable stacking fault energies (USFEs). These crystals are readily analyzable by the real-space 3D phase-field dislocation dynamics (PFDD) technique, which predict stress-driven pathways taken by individual dislocations. With the multiscale strategy, we show that SRO strengthening manifests in both MPEAs, with the average USFEs and critical stresses to initiate and sustain propagation of dislocations increasing with SRO above those for the ideal random solid solution. We introduce a figure of merit to measure SRO impact across compositions and annealing treatments and with it, reveal that CSRO strengthening contribution scales linearly with degree of SRO. The computations also reveal that gliding dislocations in subcritical conditions experience significant hardening. This glide hardening is strongly correlated to the statistical dispersion in the local USFE, and since CSRO tends to narrow the distribution in USFE, glide hardening decreases with degree CSRO. In studying dislocation loop expansion across stress regimes, a transition between jerky and smooth dislocation glide is identified and related to stress sensitivity of kink-pair nucleation rates of the screw character portions. Finally, analysis reveals that initially screw- and edge-oriented dislocations will become wavy in glide yet move via different mechanisms---kink-pair formation and migration vs. pinning/depinning. Their individual glide mechanisms do not change with composition, amount of SRO, glide distance, or subcritical or overdriven loading conditions. These computations explain why MoNbTi is the stronger one and has the greater strain hardening and forecasts that it is more amenable to SRO strengthening.

\section{Results}

\subsection{Moment tensor potential}
\begin{figure}[htp]
    \centering
    \includegraphics[width=0.85\linewidth]{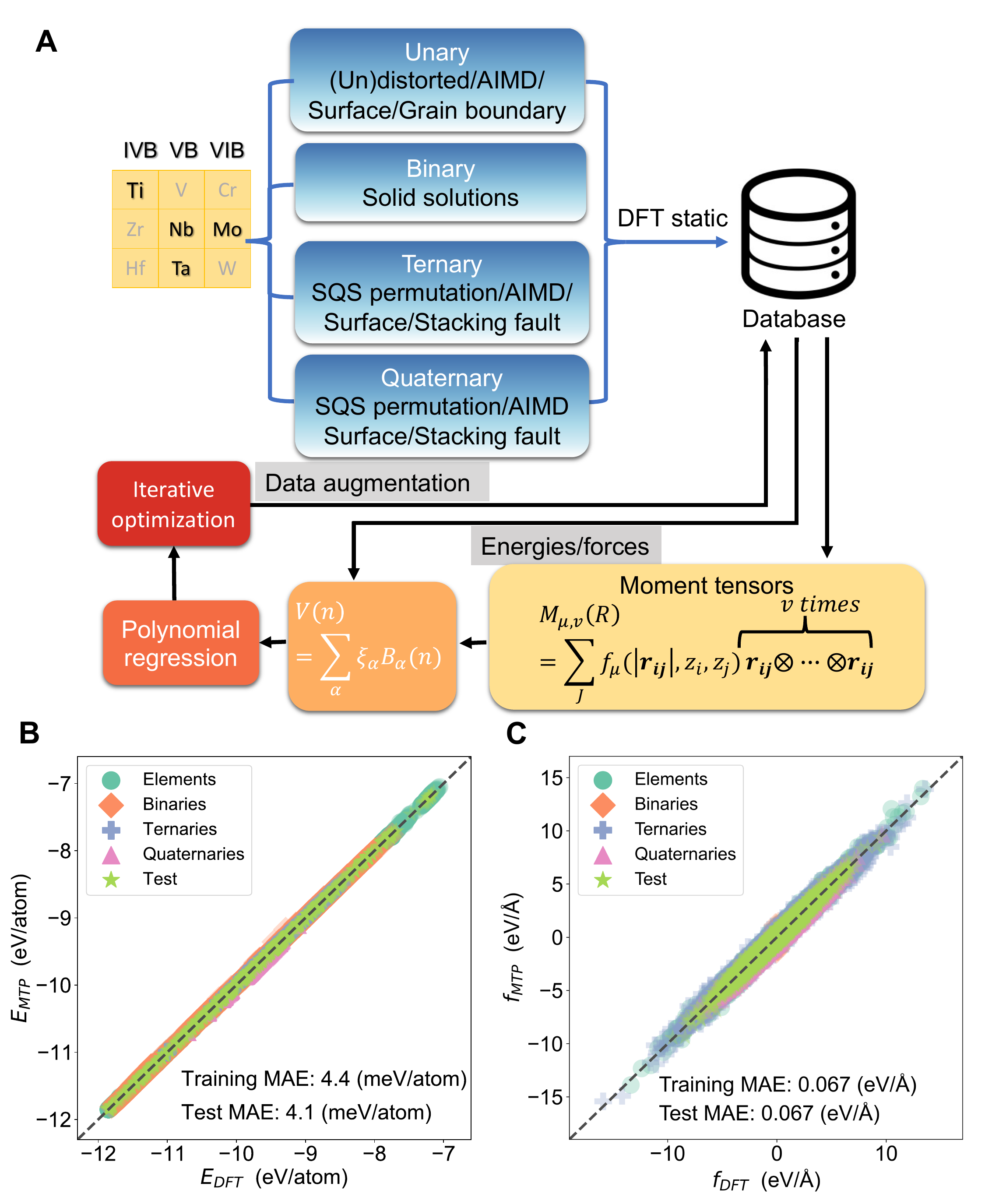}
    \caption{(a) Moment tensor potential development workflow. (b-c) Parity plots of the MTP predicted (b) energies and (c) forces against DFT values, broken down into elemental, binary, ternary and quaternary phases.}
    \label{fig:parity_plot}
\end{figure}

Studies of the RMPEAs here are enabled by the development of a highly accurate machine learning interatomic potential for the Mo-Ta-Nb-Ti system based on the MTP formalism \cite{chenAccurateForceField2017b, liQuantumaccurateSpectralNeighbor2018b,zuoPerformanceCostAssessment2020c, qiBridgingGapSimulated2021}. Fig. \ref{fig:parity_plot}a provides an overview of the MTP fitting procedure, which is based on a well-established workflow developed by some of the current authors in previous works \cite{liComplexStrengtheningMechanisms2020a, yinAtomisticSimulationsDislocation2021}. To best investigate the effect of composition variations on SRO and dislocation glide, the training data were carefully selected to encompass all known unary, binary, ternary and quaternary phases in the Mo-Ta-Nb-Ti system (see Methods for details). Fig. \ref{fig:parity_plot}(b,c) show that extremely low test mean absolute errors (MAEs) were achieved for energies (4.1 meV$\cdot$atom$^{-1}$) and forces (0.067 eV$\cdot$\AA$^{-1}$), comparable to that achieved previously for the NbMoTaW RMPEA \cite{liComplexStrengtheningMechanisms2020a, yinAtomisticSimulationsDislocation2021}. The MTP also reproduces very well the DFT elastic constants for the constituent elemental systems, as shown in Fig. \ref{SI:elastic}. The shear moduli $\mu$ are 29.6 and 32.3 GPa for TaNbTi and MoNbTi, respectively, and the Young's moduli are 82.7 and 90.7 GPa, respectively.

\subsection{Chemical short-range order}

\begin{figure}
    \centering
    \includegraphics[width=0.525\linewidth]{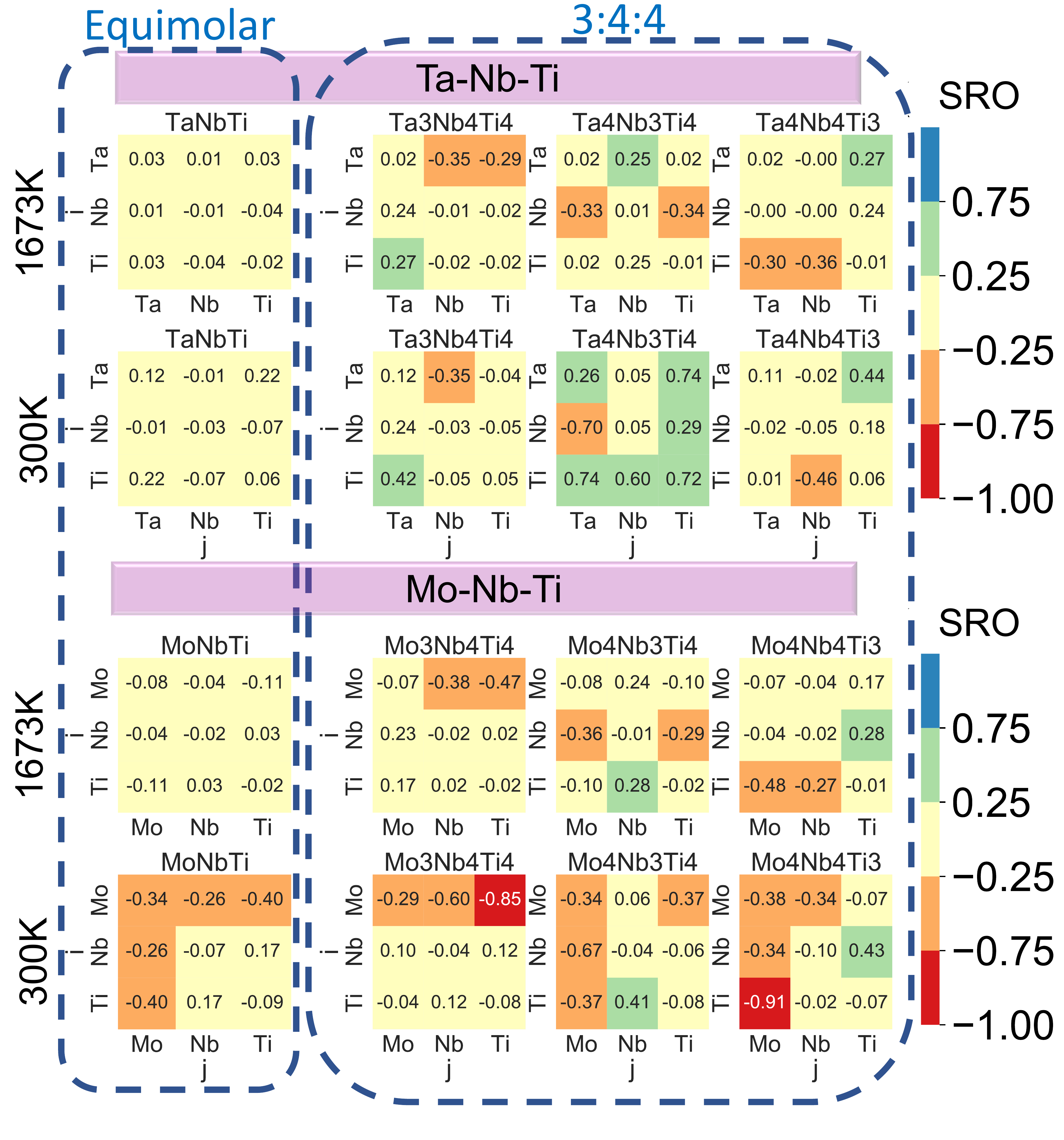}
    \includegraphics[width=0.42\linewidth]{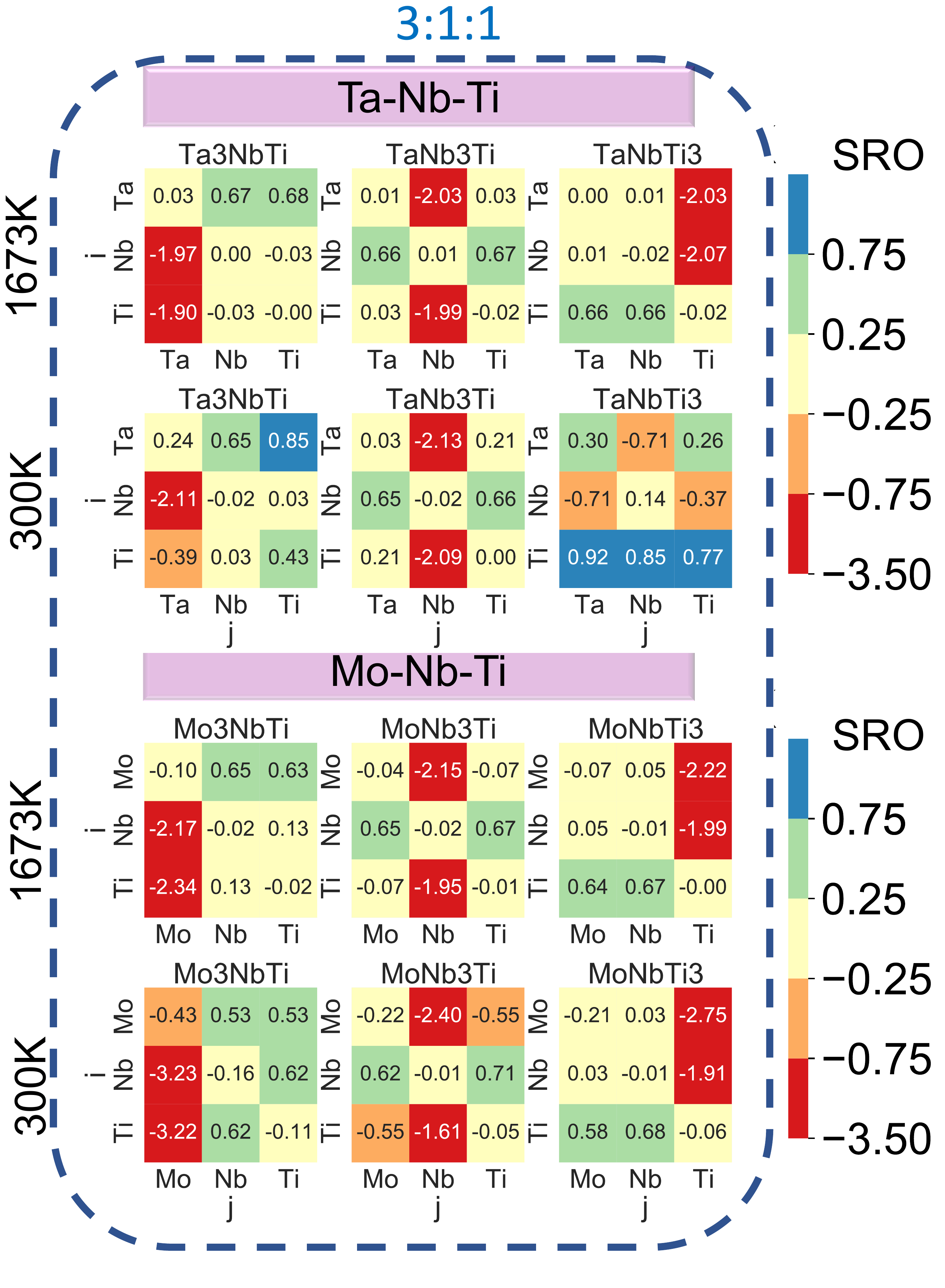}
    \caption{Heat maps of the equilibrium Warren-Cowley parameters $\alpha_{ij}$ for alloys annealed at 1673 K and 300 K. By definition, $\alpha_{ij} = \alpha_{ji}$ for equimolar systems, which is reflected in the heatmap with a diagonal symmetric color matrix. However, for the systems with non-equimolar composition, $\alpha_{ij} \neq \alpha_{ji}$. The color scale distinguishes between low SRO ($|\alpha_{ij}| < 0.25$), medium SRO ($0.25 \leq |\alpha_{ij}| < 0.75$) and high SRO ($|\alpha_{ij}| \geq 0.75$). }
    \label{fig:wc}
\end{figure}

To calculate the chemical SRO, bulk BCC supercells with equimolar MoNbTi and TaNbTi as well as non-equimolar ternaries with elemental ratio of 3:4:4 and 3:1:1, i.e., \ce{X4Nb3Ti4}, \ce{X4Nb4Ti3}, \ce{X3Nb4Ti4},  \ce{X3NbTi}, \ce{XNbTi3}, \ce{XNb3Ti}, where X = Mo or Ta, were constructed. For each composition, three levels of SRO are achieved by studying the as-constructed random solid solution (RSS) and annealing at 300 K and 1673 K using MC/MD simulations with the MTP. The SRO for the final equilibrium structures is characterized using Warren-Cowley parameters, which are defined for each pair type $ij$ as $\alpha_{ij} = (p_{ij}-c_j)/(\delta_{ij}-c_j)$, \noindent where $p_{ij}$ is the probability of having an atom type $j$ in the nearest neighbor shell of an atom type $i$ and $c_j$ is the concentration of type $j$ and $\delta_{ij}$ is the Kronecker delta function \cite{cowleyApproximateTheoryOrder1950}. By definition, for a RSS, $\alpha_{ij}\approx 0$ and for greater degrees of SRO, the absolute value of $\alpha_{ij}$ increases. Fig. \ref{fig:wc} shows the Warren-Cowley parameters for the annealed RMPEAs. For all compositions, greater levels of SRO can be achieved with the lower annealing temperature. For the same set of elements, greater degrees of SRO can be accomplished with off-equimolar stoichiometry, i.e., the 3:1:1 and 3:4:4 compositions, than equimolar. In materials annealed at 300 K, the SRO exhibited by RMPEAs containing Mo (\ce{Mo_$x$Nb_$y$Ti_$z$}) are much higher than those containing Ta (\ce{Ta_$x$Nb_$y$Ti_$z$}). These two types of RMPEAs would be expected to respond differently to the same processing condition or heat treatment, with MoNbTi being much more susceptible than TaNbTi.  



\subsection{Unstable stacking fault energy}

\begin{figure}
    \centering
    \includegraphics[width=0.85\linewidth]{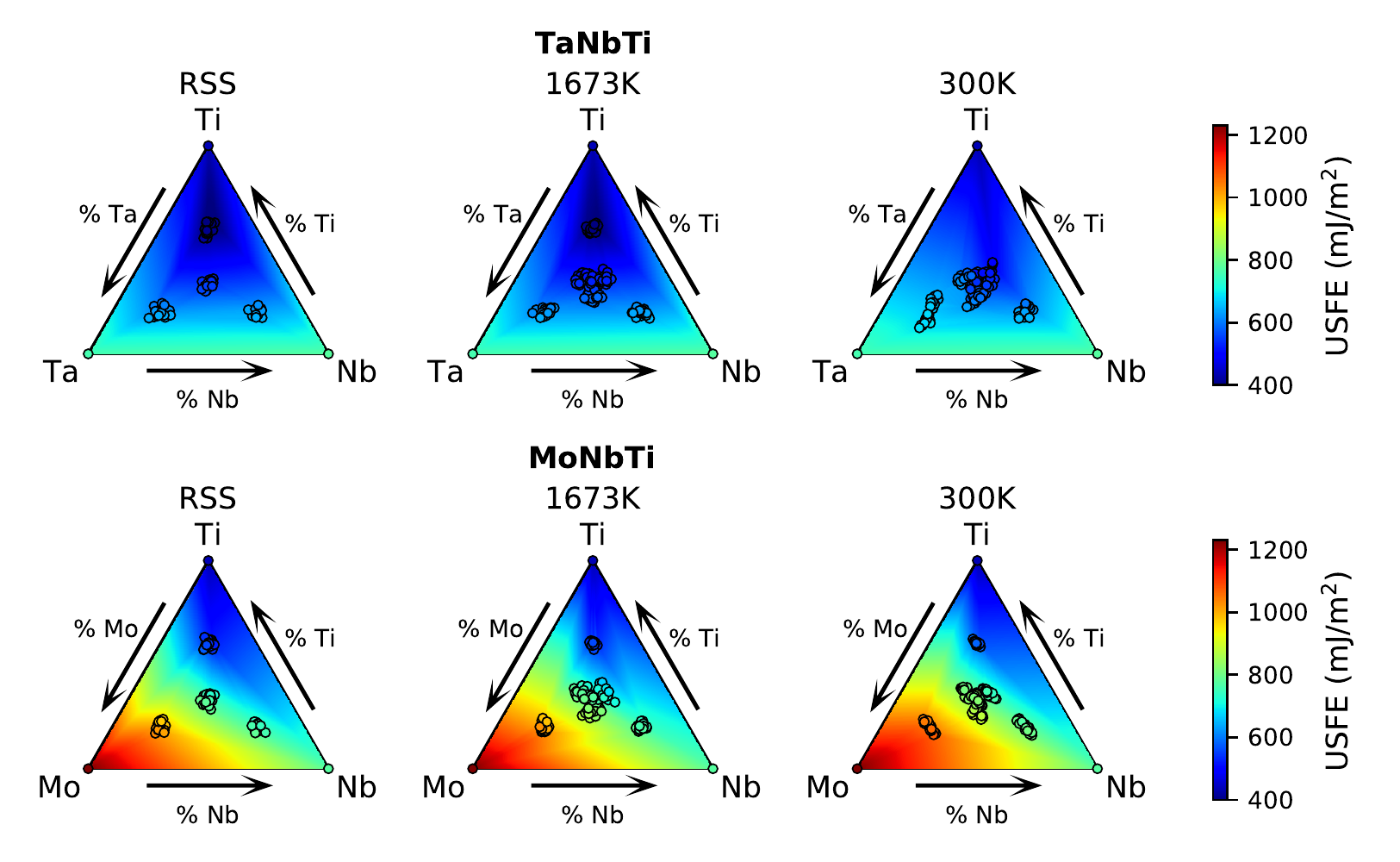}
    \caption{The USFE values calculated with the MPT as a function of the local composition around the fault plane in the two alloys at different levels of SRO. The values obtained from energy minimization using MTP are shown with dots. The values in the remainder of the triangles are interpolated from the calculated values and colored.}
    \label{fig:usfe_interp}
\end{figure}

Fig. \ref{fig:usfe_interp} plots the calculated unstable stacking fault energy (USFE) on the \{110\} plane, shifting along with $\left<111\right>$ directions for as-constructed RSS and samples equilibrated at two different annealing temperatures. Fig. \ref{SI:SFE_dist} A-D plots the USFE distributions for all the RMPEAs with different compositions. In all cases, a higher concentration of Ti reduces the USFE. In the Mo-Nb-Ti system, greater concentrations of Mo increases the USFE, in agreement with a prior work using another interatomic potential \cite{xu_atomistic_2020}. On average, for the same composition, annealing at 300 K raises the USFE  indicating that SRO increases USFE. To correlate the USFE with its local composition, the local composition is determined based on the composition of the first nearest neighbor planes surrounding the cleaving plane. The relationships between USFE and the local composition are plotted in Fig. \ref{SI:SFE_dist} E. The correlations between the USFE of a plane and its local composition are consistent with those observed for average USFE of different bulk concentrations. Higher local fractions of Mo significantly increase the USFE, while higher fractions of Ta also increase the USFE but not as significantly as Mo. Higher fractions of Ti significantly decrease the USFE. The trend of USFE with local composition is consistent with the trends observed for the bulk composition as discussed above.

\subsection{Phase-field framework}

PFDD was employed to simulate the behavior of dislocations in both alloys (see Methods, \cite{beyerleinUnderstandingDislocationMechanics2016}). All inputs into PFDD are physical, such as the elastic constants, lattice parameters, and USFE, and are obtained from the MTP calculations in the preceding sections. In an MPEA, the resistance to dislocation glide is not homogeneous within the material due to variations in local composition \cite{raoModelingSolutionHardening2019b,marescaTheoryScrewDislocation2020a, wangMultiplicityDislocationPathways2020,xuLocalSlipResistances2021,romeroAtomisticSimulationsLocal2022}. A dislocation will only be affected by the type and arrangement of atomic elements within a radius of a few Burgers vectors \cite{ghafarollahiSoluteScrewDislocation2019}. As such, at any position and point in time, the dislocation only samples a small nanoscale region of the crystal with a local composition differing from that of the bulk. In PFDD, the local composition of every nanometer region is associated with its corresponding USFE. We account for these spatial variation in composition by creating a crystal with spatial variation  USFE, simulating the variable resistance to dislocation glide \cite{smithEffectLocalChemical2020,fey_transitions_2022}.


To generate position-dependent USFE crystals for MoNbTi and TaNbTi, random lattices with target SRO levels were created using a binary alloy swapping method adapted from Gehlen et al \cite{gehlenComputerSimulationStructure1965} (see the Methods section). Each lattice point in a random lattice is then assigned a local composition, which is defined as the composition of the atom and its neighbors up to the 5th nearest neighbor within two adjacent (110) planes. The 5th nearest neighbor was chosen based on atomistic simulations of solute-dislocation interaction energies \cite{ghafarollahiSoluteScrewDislocation2019}, and only atoms within the two (110) planes that would be sheared by a dislocation were considered. A local USFE value was then determined using the USFE-composition maps in Fig. \ref{fig:usfe_interp}. 
In total, 420 lattices containing several million atoms were created for each level of SRO for both alloys, with examples shown in Fig. \ref{SI:usfe_surf}. To quantify SRO based on the Warren-Cowley parameters, we introduce $\omega$ as the Euclidean norm of the three like-pair Warren-Cowley parameters $\omega = \sqrt{\alpha_{11}^2+\alpha_{22}^2+\alpha_{33}^2}$ as a single figure of merit (FOM) $\omega$ that can be used to compare the relative extent of SRO in an equimolar alloy. With it, we find that the mean USFE scales directly with $\omega$ (Fig. \ref{SI:usfe_fom}A). The MoNbTi alloy reaches a higher $\omega$ for the same set of annealing treatment due to the preference for Mo-Ti and Mo-Nb bonds. The coefficient of variation (CV) in USFE tends to decrease with $\omega$, a consequence of the reduction in random atomic pairings (Fig. \ref{SI:usfe_fom}B).


\subsection{Dislocation glide mechanisms}

We first study the role of SRO on the glide behavior of initially straight edge or screw dislocations. Due to the randomness in underlying fault energies, twenty independent realizations are performed for each alloy and each level of SRO. To study critical behavior, the applied shear stress is gradually increased in increments of 0.001$\mu$ until the dislocation glides and is held constant until it fully arrests. 

\begin{figure}
    \centering
    \includegraphics[width=0.8\linewidth]{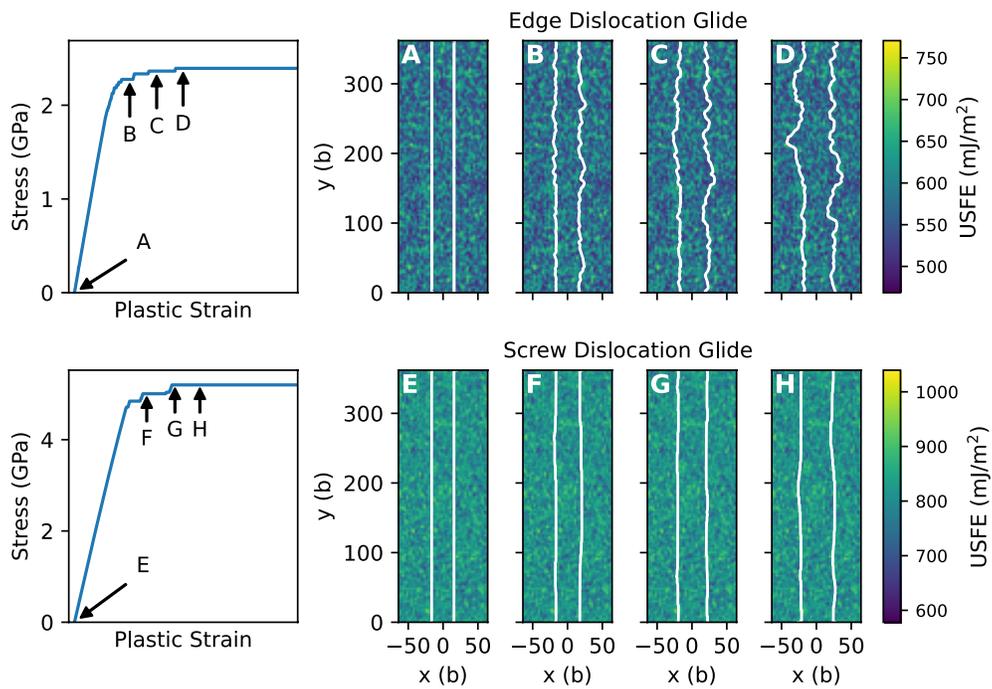}
    \caption{Representative examples of edge (A-D) and screw (E-H) dislocation glide and their associated stress-strain curve. The edge example is in TaNbTi at 300 K SRO, and screw example is in MoNbTi at 300 K SRO.  }
    \label{fig:glide}
\end{figure}

Fig. \ref{fig:glide} shows snapshots in time of edge dislocation glide. When the stress is initially applied and raised, the dislocation remains straight. Once the applied stress exceeds the first threshold, the edge dislocation becomes slightly wavy, as small portions of the dislocation line bow out into low USFE regions and are held back at the higher USFE regions. The stress must be increased further for the dislocation to glide, and small bowed out segments of the dislocation will glide independently through the lower USFE regions, dragging the neighboring (non-edge) segments through the higher USFE regions. The dislocation arrests several times during the simulation, each time requiring the stress to be raised to restart glide. The arrested dislocation morphologies are wavy, unlike the original pure edge orientation. The stop/start behavior leads to glide plane hardening, a continual increase in applied stress with increasing plastic strain, as seen in the stress-strain curve in Fig. \ref{fig:glide}. This is in contrast to PFDD simulations of pure metals, in which dislocations gliding have a single critical stress and remain straight during glide. 

Screw dislocation glide proceeds in a different manner from edge dislocation glide. As the stress increases, the dislocation remains completely straight with pure screw character until a kink-pair only a few Burgers vectors wide is nucleated into a low USFE region. Unlike the variable wavy bow out in the edge dislocation, these kink-pairs always have a height of just 1$b$, and the kinks will usually, but not always, glide along the length of the screw dislocation to advance the full dislocation line forward. Like the edge dislocations, the screw dislocations may become arrested under stress, but unlike the edge dislocations, the arrested dislocation morphologies are nearly straight, apart from a few metastable kinks, recovering the original pure screw orientation.  The start/stop mechanism of glide of the screw dislocation also leads to glide plane hardening, although at a lower level than edge dislocation glide plane hardening.

\subsection{Dislocation glide stresses}

\begin{figure}
    \centering
    \includegraphics[width=0.85\linewidth]{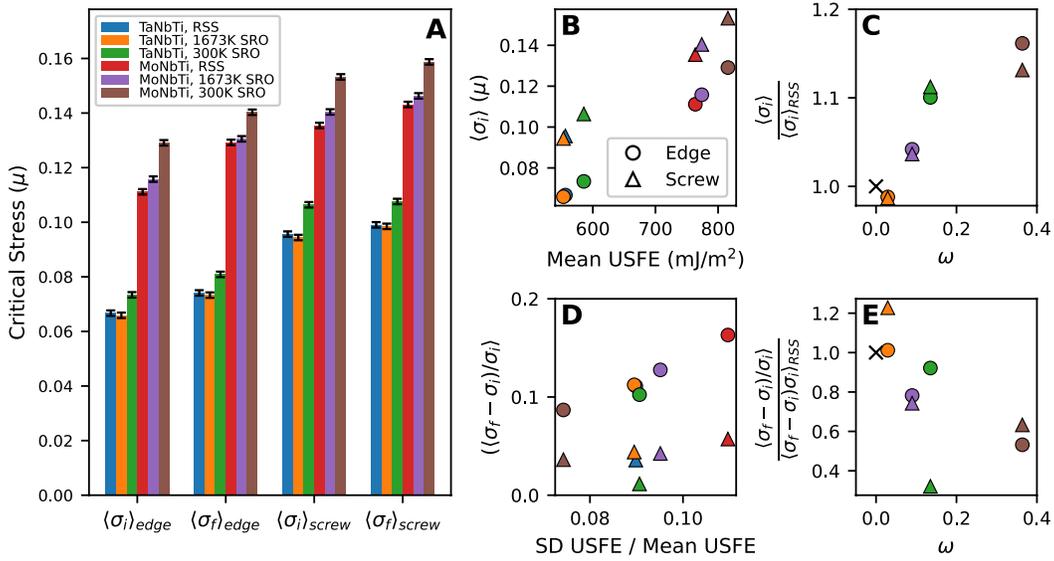}
    \caption{Critical stresses for dislocation glide. (A) The mean critical stress for dislocations in each of the alloys. The critical stresses for MoNbTi and TaNbTi are normalized by their respective shear moduli $\mu$. The error bars show the resolution of the simulation, 0.001 $\mu$. (B) The mean initial critical stress vs the mean USFE for each alloy. (C) The mean initial critical stress relative to the mean initial critical stress for RSS alloy vs the extent of SRO, represented by the SRO FOM. The black $\times$ represents the RSS case. (D) The hardening, represented by the difference in final and initial critical stresses, vs the coefficient of variation of USFE. (E) The hardening relative to the hardening for the RSS alloy vs the extent of SRO.  }
    \label{fig:crit_stress}
\end{figure}

Fig. \ref{fig:crit_stress}A plots the stresses to initiate glide $\sigma_i$ and the final stresses for runaway glide $\sigma_f$ of edge and screw dislocations based on twenty independent initializations. Regardless of SRO, both $\sigma_i$ and $\sigma_f$ for the TaNbTi alloy are lower than those for the MoNbTi alloy. From Fig. \ref{fig:crit_stress}B, we find that mean glide resistance across the plane scales directly with the average USFE across the plane. Thus, changes in USFE caused by SRO have a direct influence on the stress to initiate and propagate dislocations, as seen in Fig. \ref{fig:crit_stress}C. The higher the $\omega$, the higher the USFE is increased relative to the RSS case, which translates directly to increased glide resistance for both screw and edge dislocations. 

We also find that the hardening in glide resistance is related to the degree of dispersion of the USFE values in the glide plane as opposed to the mean. In Fig. \ref{fig:crit_stress}D, we analyze the role of composition and its fluctuations by adopting the fractional increase from $\sigma_i$ to $\sigma_f$ as a measure of glide-plane hardening. Importantly, for both alloys and screw/edge character dislocations, hardening scales directly with the USFE CV. The strikingly linear relationship even when considering both alloys implies that it transcends composition. Thus, apparent differences in the hardening seen in these alloys can be explained. Compared to TaNbTi, MoNbTi achieves, on average, greater hardening in the ideal random case and lower hardening in the highest $\omega$ case.

Further, while hardening for both screw and edge dislocations increase as the USFE CV increases, the edge dislocations experience greater hardening than the screw dislocations for the same statistically sampled glide plane length (Fig. \ref{fig:crit_stress}D). The edge dislocations glide by depinning of the segments at the relatively harder regions, segments which have reoriented to non-edge character due to bow out. Continued glide, therefore, relies on overcoming those local regions of higher resistance. Encountering a region ahead of the dislocation of even greater resistance than in the wake more likely occur when the dispersion in USFE is greater. The screw dislocation moves by producing short and narrow atomic advances of screw-oriented segments, i.e., kink-pairs, in the weaker regions and relying on the long advances of easier-to-move edge segments along the length of the dislocation. Those local domains of higher resistance that are more likely encountered when the dispersion in USFE is higher, can be easily overcome by migrating edge dislocations. By virtue of their differing glide mechanisms, edge dislocations experience greater sensitivity to the dispersion in USFE and hence greater hardening than screw dislocations. 

Fig. \ref{fig:crit_stress}E examines the influence of $\omega$ on hardening. It reveals that the role of SRO on hardening corresponds to the extent to which SRO affects the CV in USFE.  As increased $\omega$ tends to narrow the dispersion in USFE across the glide plane, it reduces glide-plane hardening.  Since the TaNbTi alloy achieves lower $\omega$ than MoNbTi for the same annealing treatment, hardening, like its strength, is weakly affected by SRO compared to MoNbTi. 

\subsection{Effect of chemical fluctuations on glide mobility}

\begin{figure}
    \centering
    \includegraphics[width=0.8\linewidth]{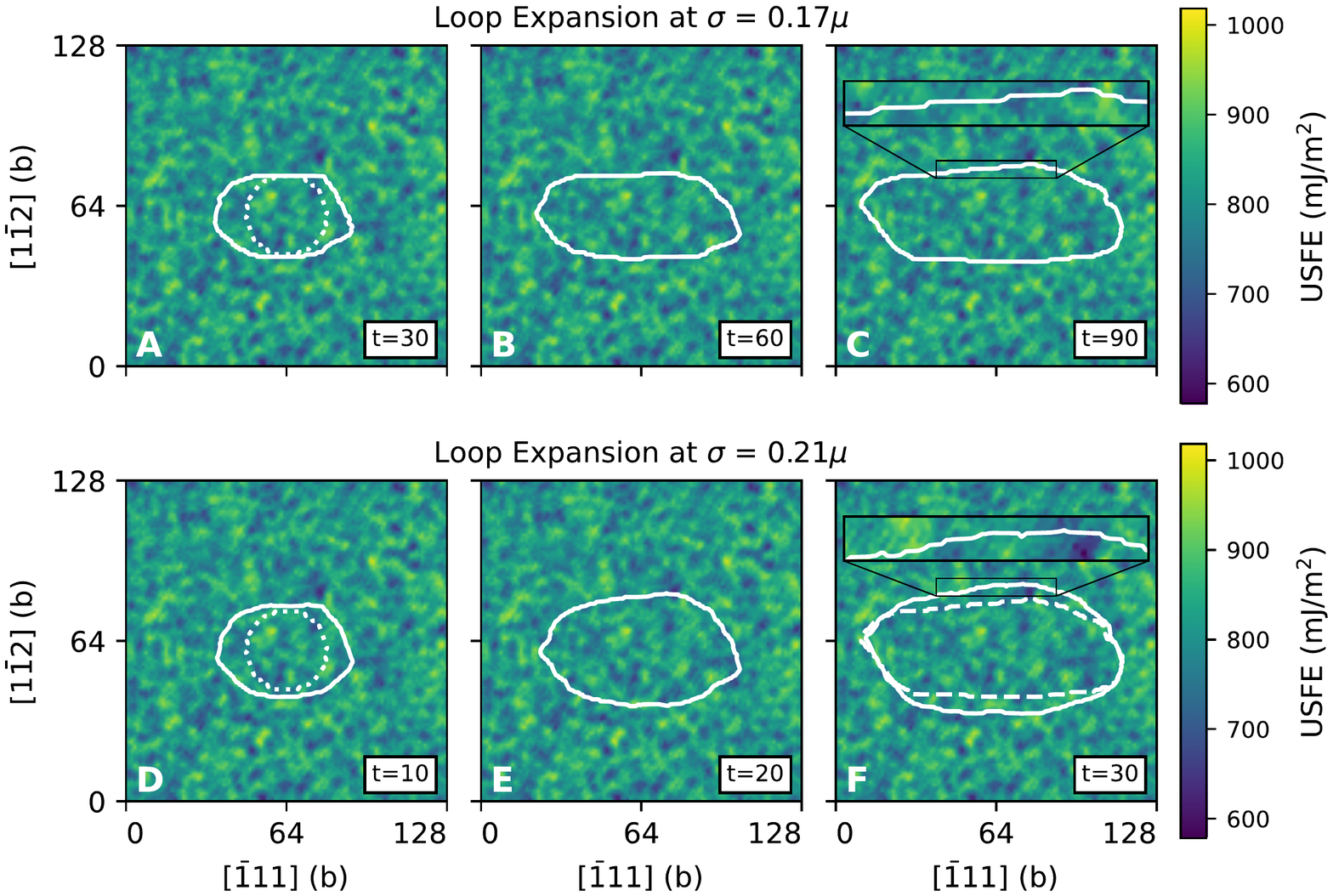}
    \caption{The same dislocation loop expanding in MoNbTi under different applied stresses. The initial loop shape is shown by the dotted lines in (A) and (D). When a lower stress is applied, the screw dislocation nucleates kink-pairs infrequently, causing jerky dislocation glide and remaining largely pure screw. At higher applied stresses, the screw dislocation nucleates many kink-pairs at once resulting in smoother glide and a wavy morphology. The final loop from (C) is reproduced by the dashed line in (F) to highlight the difference in aspect ratio between the two loops.}
    \label{fig:loops}
\end{figure}

Next, we study the effect of chemical fluctuations on screw/edge glide mobility by examining loop expansion on the $(110)$ slip plane at constant stress. For given applied stress, thirty dislocation loop expansion simulations are conducted representing different locations in a given alloy and SRO. We find that for all cases, the anisotropy in screw/edge behavior reduces with increases applied stress. At low stresses, the difference in screw/edge behavior is large, causing the loop to expand into an oblong shape (Fig. \ref{fig:loops}A-D). The edge segments move continuously, constantly changing their wavy appearance. The screw dislocations advance very slowly, nucleating only a few kink-pairs at a time and recovering the nearly straight orientation with each advancement. As the applied stress increases, both screw and edge velocities increase and their ratio decreases towards unity. The loop expands more isotropically, almost FCC-like (Fig. \ref{fig:loops}E-H). Figure \ref{fig:loops}H shows the loop at both low and high stress overlaid together. The two loops have similar widths (103$b$ and 101$b$, respectively) but different heights (38$b$ and 51$b$, respectively), demonstrating that higher stresses decrease the loop aspect ratio. The screw dislocation has clearly changed its mode of glide, as a result of a higher kink-pair nucleation rate. The screw portions move continuously and adopt a wavy appearance, superficially much like the edge dislocations. Wavy screw glide, however, occurs as many kink-pairs nucleate simultaneously along the same dislocation. Newly advanced portions can nucleate further kink pairs, causing different portions of the dislocations to advance at different rates.

\begin{figure}
    \centering
    \includegraphics[width=0.85\linewidth]{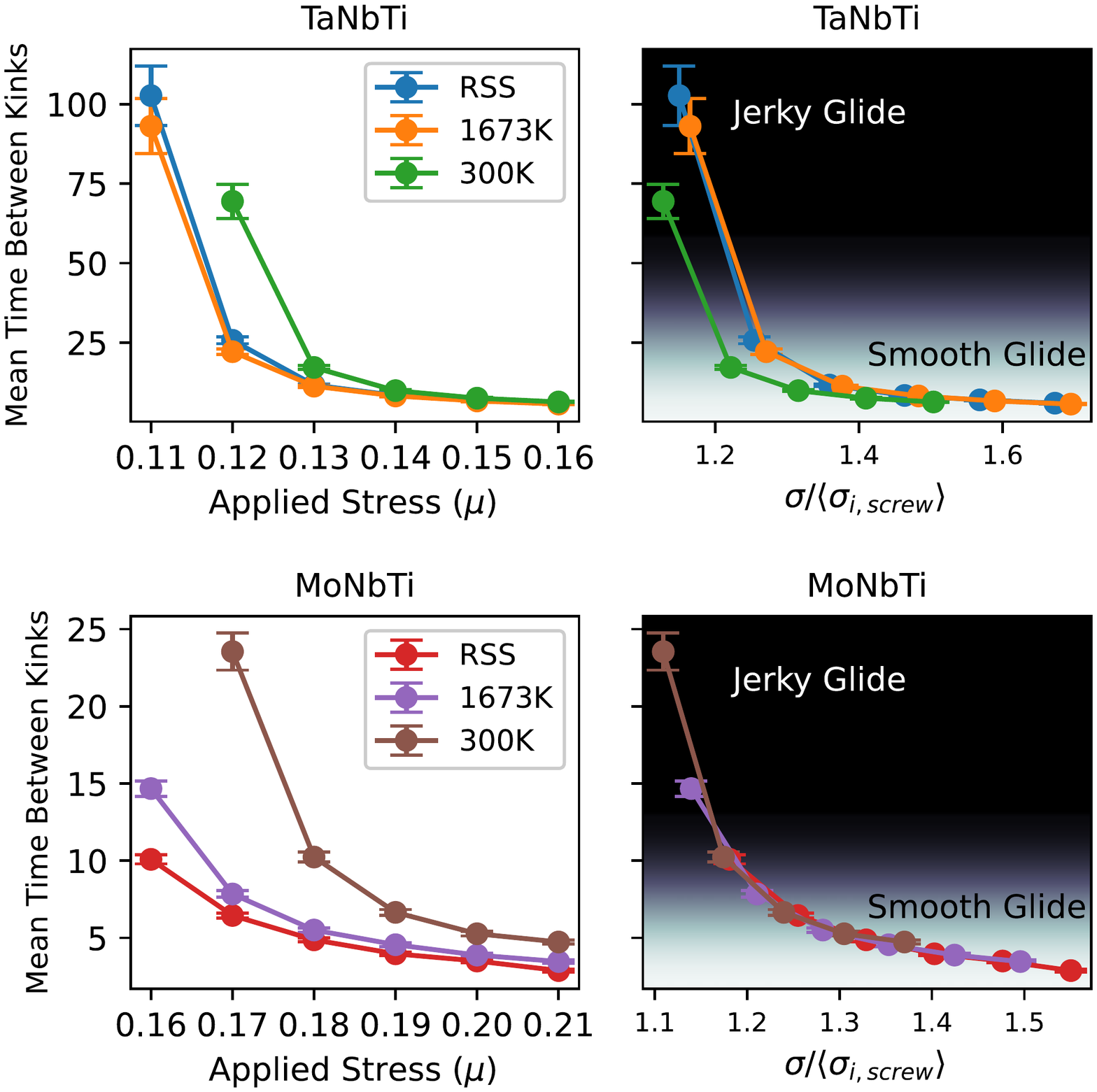}
    \caption{Average time between kink-pair nucleation events as a function of applied stress. The error bars show the standard error. The transition between jerky and smooth screw dislocation glide is determined by the applied stress relative to the dislocation critical stress.}
    \label{fig:kink_times}
\end{figure}

We separate the two extremes of screw dislocation behavior into ``jerky" glide at low stresses and ``smooth" glide at high stresses. To link the transition between jerky and smooth dislocation the rate of kink-pair nucleation, we calculate the waiting time between kink-pair nucleation events for all simulations. The waiting times from all thirty instantiations of loops are combined into a single distribution for a given alloy and applied stress, and the means are plotted in Fig. \ref{fig:kink_times}. Over 22,000 and 52,000 total waiting times were recorded for TaNbTi and MoNbTi, respectively, and each individual distribution contains at least 200 values. For both TaNbTi and MoNbTi, higher levels of SRO correspond to longer average waiting times and thus more jerky dislocation glide at all applied stresses. Normalizing the applied stresses by the average $\sigma_i$ for screw glide, the three distinct SRO curves collapse into one. Thus, the critical stress to transition from jerky to smooth scales with Peierls strength or with static strength. SRO affects the transition stress in dynamic glide indirectly via its strengthening effect on static glide resistance.  

\section{Discussion}

There are limited experimental measurements of the mechanical properties of TaNbTi and MoNbTi. The tensile yield stresses have been reported as 620 and 950 MPa for TaNbTi and MoNbTi, respectively  \cite{zykaMicrostructureRoomTemperature2019,wangMultiplicityDislocationPathways2020,senkovCorrelationsImproveRoom2021}, which are consistent with the dislocation dynamics predictions of higher glide stresses for MoNbTi. The ultimate tensile strengths are 683 and 1500 MPa, respectively, so MoNbTi exhibits significant strain hardening while TaNbTi does not. Our simulations show higher hardening in MoNbTi than TaNbTi due to the increased coefficient of variation in USFE. The dislocation dynamics simulations also revealed that the amount of dislocation hardening is decreased by the presence of SRO, especially for MoNbTi. As edge dislocations undergo more hardening than screw dislocations, we predict the differences in macro-scale strain hardening in these alloys are largely controlled by edge dislocation behavior. 

From the Warren-Cowley parameter calculations, it is clear that MoNbTi has a higher propensity for SRO and will be more affected by processing conditions. There has been interest in tuning the SRO parameters through heat treatment, although experimentally this is difficult to achieve \cite{dingTunableStackingFault2018d}. Findings indicate that MoNbTi is a better candidate for exploring SRO strengthening than TaNbTi since SRO promotes two relatively stronger Mo-Nb and Mo-Ti bonds. However, the relative increases in the average dislocation glide resistance due to SRO amount to less than 20\% even in the most extreme cases, so large changes in the mechanical properties must be accompanied by changes in the chemical composition, not SRO alone. 

Via MD simulations, a few studies have shown SRO-enhanced dislocation glide resistance  \cite{liStrengtheningMultiprincipalElement2019c,dingTunableStackingFault2018d,yinAtomisticSimulationsDislocation2021} and one study on CoFeNiTi alloy reported a slight SRO softening  \cite{antillonChemicalShortRange2021}. Strengthening or softening was related to the formation of immobile dislocation segments via cross slip. Here, we demonstrate SRO strengthening in dislocation glide without cross slip. While the current dislocation model permits cross slip \cite{feyPhasefieldDislocationModeling2022}, it was not observed in the present calculations since the influence of thermal fluctuations is not taken into account. Including temperature would undoubtedly increase the chance for cross-slip or cross-kinking, adding another mechanism for SRO strengthening. 

\section{Methods} \label{Methods}

\subsection{Machine-learning interatomic potential}

MTP, which has been shown to yield an excellent balance between accuracy and computational cost \cite{zuoPerformanceCostAssessment2020c}, was utilized in this work. Details about the MTP formalism can be found in previous works for interested readers \cite{shapeevMomentTensorPotentials2016b, gubaevAcceleratingHighthroughputSearches2019, podryabinkinAcceleratingCrystalStructure2019}. Briefly, in MTP model, the energy $E$ is partitioned into the contributions ($V$) of individual atomic neighborhood ($\boldsymbol{n_i}$). $V(\boldsymbol{n})$ is linearly expanded through a set of basis functions $B_{\alpha}$:

\begin{equation}
    V(\boldsymbol{n}) = \sum_{\alpha} \xi_{\alpha} \boldsymbol{B_{\alpha}(\boldsymbol{n})}
\end{equation}

where the basis functions $\boldsymbol{B}_{\alpha}(\boldsymbol{n})$ are given as:

\begin{equation}
        M_{\mu, \nu}(\boldsymbol{n_i}) = \sum_{j} f_{\mu}(|\boldsymbol{r_{ij}}|,z_i, z_j)\ \underbrace{\boldsymbol{r}_{ij} \otimes \dots \otimes \boldsymbol{r}_{ij}}_{\nu\ times},    
\end{equation}

\noindent where $\boldsymbol{r_{ij}}$ is the position of the neighbor $j$th atom relative to the $i$th atom, and $f_{\mu}$ are the radial functions, which only depend on the interatomic distance $|\boldsymbol{r_{ij}}|$ and atomic types $z_i$ and $z_j$. The terms $\boldsymbol{r}_{ij} \otimes \dots \otimes \boldsymbol{r}_{ij}$ include angular information about the neighborhood $\boldsymbol{n_i}$ and are tensors of rank $\nu$. The summation is carried out over all the atoms in the neighborhood $\boldsymbol{n_j}$ around $i$th atom. $M_{\mu, \nu}$, i.e., the moment tensors, are invariant with respect to translation of the system and permutation of equivalent atoms.

The cutoff distance $R_{cut}$ and maximum level $lev_{max}$ are the two important hyper-parameters that determine the accuracy and computational cost of MTP. Their values in this work are set to 4.8 \r{A} and 20, respectively. The loss function used for optimizing the MTP is given as:

\begin{equation}
    L(\boldsymbol{\theta}) = \sum_j w_e(E(\boldsymbol{\theta}, x^{{j}})- E^{qm}(x^{(j)}))^2 + \sum_jw_f(\boldsymbol{f}(\boldsymbol{\theta}, x^{{j}})-\boldsymbol{f}^{qm}(x^{(j)}))^2
\end{equation}
\noindent where $\boldsymbol{\theta}$ are the set of parameters to be optimized; 
$x^{{j}}$ are the configurations in the training data; $E(\boldsymbol{\theta}, x^{{j}})$ is the MTP predicted energy; $E^{DFT}(x^{(j)})$ is DFT energy; $\boldsymbol{f}(\boldsymbol{\theta}, x^{{j}})$  is the MTP predicted force; and $\boldsymbol{f}^{qm}(x^{(j)})$ is DFT force. $w_e$, $w_f$ are non-negative weights expressing the importance of energies, forces in the optimization problem.

\subsection{Training data generation}
The comprehensive training data are essential to the robust potential development. The training data include the elemental, binary, ternary, and quaternary chemical systems. From the structure type perspective, ground state bulk structures, distorted bulk structure, surfaces, stacking faults, and grain boundaries are also considered to improve the accuracy for potential in calculating planar defects. The detailed structure generation for each system is provided as follows.
 \begin{enumerate}
    \item Elemental systems (BCC Mo, Ta, Nb, Ti and HCP Ti)
    \begin{enumerate}
        \item Ground state structures for the elements, i.e., BCC Mo, Ta, Nb, and HCP Ti, plus BCC Ti;
        \item Distorted structures constructed by applying strains of $-10\%$ to 10$\%$ at 1$\%$ intervals to the bulk conventional cell for BCC Mo, Ta, Nb, Ti and HCP Ti. Detailed can be found in ref. \cite{dejongChartingCompleteElastic2015c};
        \item Surface structures of elemental metals obtained from the Crystalium database \cite{tranSurfaceEnergiesElemental2016c};
        \item Grain boundary structures of elemental metals obtained from grain boundary database \cite{zhengGrainBoundaryProperties2020c}; 
        \item Snapshots from $NVT$ \textit{ab initio} molecular dynamics (AIMD) simulations of the bulk $3 \times 3 \times 3$ supercell for BCC and $4 \times 4 \times 2$ supercell for HCP Ti at room temperature, medium temperature (below melting point), high temperature (above melting point). In addition, snapshots were also obtained from $NVT$ AIMD simulations at room temperature at 90\% and 110\% of the equilibrium 0 K volume. Forty snapshots were extracted from each AIMD simulation at intervals of 0.1 $ps$; 100 snapshots from  high-temperature (above melting point) $NPT$ simulations are also included for each element.  
    \end{enumerate}
    \item Binary systems (Mo-Ta, Mo-Nb, Mo-Ti, Ta-Nb, Ta-Ti, Nb-Ti)
    \begin{enumerate}
        \item Solid solution structures constructed by partial substitution of $2 \times 2 \times 2$ bulk supercells of one element with the other element. Compositions of the form \ce{A_$x$B_$1-x$} were generated with $x$ ranging from 0 at\% to 100 at\% at intervals of 6.25 at\%.
    \end{enumerate}
    \item Ternary and quaternary systems (Mo-Ta-Nb, Mo-Ta-Ti, Mo-Nb-Ti, Ta-Nb-Ti, Mo-Ta-Nb-Ti)
    \begin{enumerate}
        \item Special quasi-random structures (SQS) \cite{zungerSpecialQuasirandomStructures1990c} generated with ATAT \cite{vandewalleAlloyTheoreticAutomated2002b} using a $4 \times 4 \times 4$ bcc supercell. 
        \item Snapshots from $NVT$ AIMD simulations of the Mo-Nb-Ti SQS structure at 300, 1000, 3000 K with addition of 100 snapshots structures from $NPT$ AIMD simulations at 3200 K. Snapshots from $NPT$ AIMD simulations of Ta-Nb-Ti SQS at 300, 900, 1500, 2400, 3000 K. 
        \item Surface structures \cite{tranSurfaceEnergiesElemental2016c}, stacking fault structures samples \cite{xu_atomistic_2020, huScreeningGeneralizedStacking2021} for Mo-Nb-Ti, Ta-Nb-Ti, and Mo-Ta-Nb-Ti systems. 
        \item Stacking fault structures obtained from Refs. \cite{huScreeningGeneralizedStacking2021, xu_atomistic_2020} for ternary and quaternary systems.
    \end{enumerate}
\end{enumerate}

  With each alloying percentage, we performed structure relaxations for all symmetrically distinct binary solid solution structures. We include both unrelaxed and relaxed structures in the data set. For ternary and quaternary systems, we considered the permutation of the elements in SQS structures. We also included structures from each ionic relaxation step of permutated SQS structure samples in the data set. 
  
  All the training data were generated using the same convergence criteria used in our previous work \cite{liComplexStrengtheningMechanisms2020a}. The database includes 17210 static calculations as training data and 1660 as test dataset.


\subsection{Atomistic simulations}
All the atomistic simulations using MTP were performed using LAMMPS \cite{plimptonFastParallelAlgorithms1995d}. The simulation cells are oriented along the [112], [$\Bar{1}$10] and [11$\Bar{1}$] cubic crystollagraphic directions, which are the main directions of interest for slip in the bcc system. A supercell of dimensions of $\sim$ 48 \AA $\times$ 46 \AA $\times$ 46 \AA\ with 5760 atoms was used.

\begin{figure}
    \centering
    \includegraphics[width=0.85\linewidth]{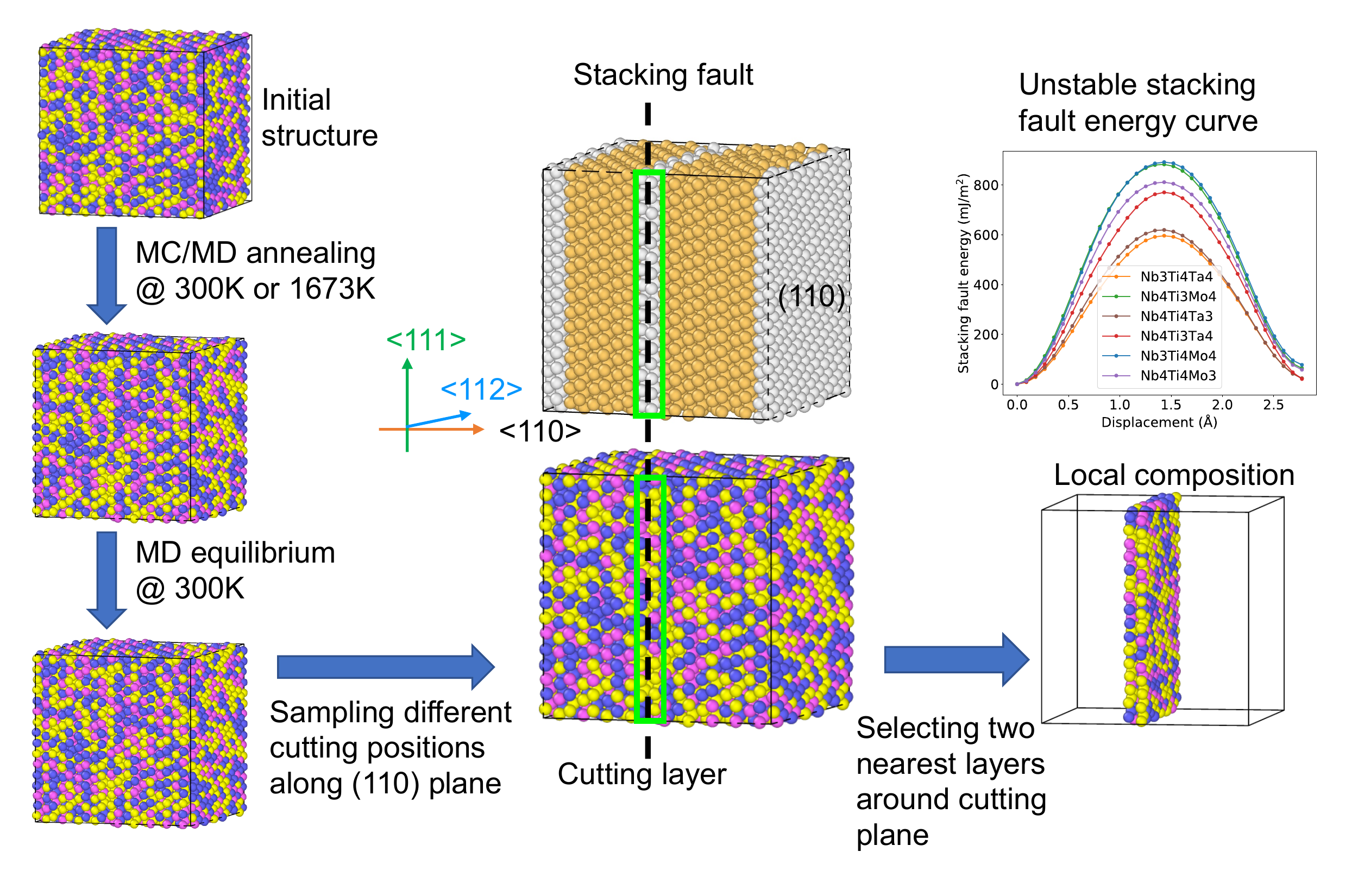}
    \caption{The procedures of sampling the stacking faults with different local compositions. }
    \label{fig:SFE_MD_samples}
\end{figure}

\begin{itemize}
    \item Hybrid MC/MD simulations within the isothermal-isobaric (NPT) ensemble were carried out for different compositions (equimolar, element ratio of 3:3:4 and 1:1:3) for ternary alloys Mo$_x$Nb$_y$Ti$_z$ and Ta$_x$Nb$_y$Ti$_z$. The supercells were randomly initialized based on the desired stoichiometry, i.e., a RSS. The simulations were then carried out at 300 K and 1673 K to obtain the equilibrium configurations with different degrees of short-range order. The Metropolis algorithm was used to attempt a swap between species type 1 and 2, species type 1 and type 3, and species type 2 and type 3 at every time step. The MD time step was set to 2 fs. The evolutions of energy, temperature, and SROs of different pairs for various compositions from MC/MD calculations can be found in SI Fig. \ref{SI:SRO_vs_step_equi} -\ref{SI:SRO_vs_step_311_MoNbTi}. After the converged configurations were reached through MC/MD, all the systems were then relaxed at 300 K using NPT MD. The structures with different composition and SROs were used for stacking fault energy calculations with energy minimization at 0 K. 
    
    \item USFE calculations. The conjugate gradient algorithm was used to perform stacking fault structure relaxation and energy minimization. The force components within (110) plane are set to zero. The direction perpendicular to (110) plane was allowed to fully relax.
    The schematic in Fig. \ref{fig:SFE_MD_samples} shows the procedures of sampling the stacking faults with different local compositions. From the equilibrated bulk MPEA structures, different stacking faults were sampled by changing the cutting layer positions along (110) plane, and shifting the right chunk of bulk along with $\left<111\right>$ direction. The composition of two atomic layers along (110) plane at the cutting position is used to represent the local composition of the sampled stacking fault.
    
\end{itemize}

\subsection{Creation of Random Lattices}

Our method to generate the random lattices with SRO is based on the method of Gehlen and Cohen, who used a swapping procedure to create binary FCC lattices with SRO \cite{gehlenComputerSimulationStructure1965}. We extend this method to multi-component alloys and only consider SRO in the first nearest neighbor (NN) shell. First, a BCC lattice of the desired shape is created, and each lattice point is randomly assigned an element type such that the overall composition is equilimolar. The total number of each bond type (e.g. Nb-Nb, Nb-Ti, etc.) is calculated. Using the Warren-Cowley SRO parameters, a goal number of each type of bond can be determined. Atoms are then swapped until the actual number of each bond type is equal to the goal. At each iteration, two lattice points with unlike element types are randomly selected. If swapping these atoms will move the bond numbers closer to the goal, the atoms are swapped. If not, the swap is rejected and a new pair of lattice points is selected. To increase the likelihood that a swap is accepted, the atoms selected for a swap are not chosen from the entire lattice but instead from a subset of atoms with desirable neighborhoods. For each element type, the expected number of each type of element in its NN shell can be predicted. When randomly selecting atoms to swap, we only choose from lattice points that do not have the expected atoms in their NN shell. For example, if the current number of $i$-$j$ bonds is higher than the goal number of $i$-$j$ bonds, we can select an $i$-atom with more $j$-atoms than expected in its NN shell to swap with a $j$-atom with more $i$-atoms than expected in its NN shell. Using this smarter swapping method for an MoNbTi lattice with the 300 K SRO parameters increases the acceptance percentage from 15\% to 63\% and decreases the running time by more than a factor of 10. 

Once the lattice has the desired SRO parameters, we assign each point a composition based on the type of the atom and up to its 5th nearest neighbor within two adjacent (110) planes for a total of 31 atoms. This composition is identified on the linearly interopolated USFE plots in Fig. \ref{fig:usfe_interp} in order to assign each lattice point a USFE value. 

\subsection{Phase-Field Dislocation Dynamics}
PFDD is a mesoscale model used to study dislocation glide through crystals. The full details of the PFDD method are given elsewhere \cite{beyerleinUnderstandingDislocationMechanics2016,xu_frank-read_2020, smithEffectLocalChemical2020}. PFDD tracks the dislocation configuration through scalar order parameters $\phi^\alpha(\mathbf{r})$ where $\alpha$ corresponds to a slip system with Burgers vector $\mathbf{b}^\alpha$ and plane normal $\mathbf{n}^\alpha$. When $\phi^\alpha(\mathbf{r})$ = 0 or 1, point $\mathbf{r}$ is unslipped or slipped by a full dislocation of type $\alpha$, respectively. Intermediate values of $\phi^\alpha(\mathbf{r})$ at the interface between slipped and unslipped regions correspond to dislocations. The order parameters are used to calculate a total system energy $\psi(\mathbf{\phi}(\mathbf{r}))$, which is then minimized by the Ginzburg-Landau equation to evolve the dislocations. 

\begin{equation}
    \frac{\partial \phi^\alpha}{\partial t} = -m_0 \frac{\partial \psi}{\partial \phi^\alpha} 
    \label{eq:GL}
\end{equation}

\noindent where $t$ is time and $m_0$ is a dislocation mobility coefficient. There are three energy contributions to the total energy: the lattice energy $\psi_{latt}(\phi(\mathbf{r}))$, the elastic energy $\psi_{elas}(\phi(\mathbf{r}))$, and the external energy $\psi_{ext}(\phi(\mathbf{r}))$:

\begin{equation}
    \psi(\mathbf{\phi}(\mathbf{r})) = \psi_{latt}(\mathbf{\phi}(\mathbf{r}))+\psi_{elas}(\mathbf{\phi}(\mathbf{r}))-\psi_{ext}(\mathbf{\phi}(\mathbf{r}))
    \label{eq:tot_energy}
\end{equation}

The elastic energy accounts for the elastic strain in the crystal and is given by 

\begin{equation}
    \psi_\mathrm{elas}(\mathbf{\phi}(\mathbf{r})) = \frac{1}{2}c_{ijkl} \epsilon_{ij}^e\left(\mathbf{\phi}(\mathbf{r})\right) \epsilon_{kl}^e\left(\mathbf{\phi}(\mathbf{r})\right)
    \label{eq:elas_energy}
\end{equation}

\noindent where $c_{ijkl}$ is the elastic stiffness tensor and $\epsilon_{ij}^e$ is the elastic strain which can be calculated from the order parameters \cite{xu_modeling_2019}. The external energy accounts for the work done by an applied stress field and is given by 

\begin{equation}
    \psi_\mathrm{ext}(\mathbf{\phi}(\mathbf{r})) = \sigma^\mathrm{app}_{ij} \epsilon^p_{ij}(\mathbf{\phi}(\mathbf{r}))
\end{equation}

\noindent where $\sigma^\mathrm{app}_{ij}$ is the applied stress state.

The final energy term, lattice energy represents the energy to break bonds at the dislocation core, and its form is specific to the material being studied. Because dislocations in BCC crystals have compact cores, a simple sinusoidal approximation is used \cite{peng3DPhaseField2020}

\begin{equation}
    \psi_{latt}(\mathbf{\phi}(\mathbf{r})) = \sum_{\alpha=1}^N \frac{\gamma_{usf}(\mathbf{r})}{d} \sin^2(\pi\phi^\alpha(\mathbf{r}))
    \label{eq:latt_energy}
\end{equation}

\noindent where $\gamma_{usf}$ is the USFE, $d$ is the slip plane interplanar spacing, and $N$ is the number of slip systems.

In each PFDD simulation, three order parameters are used to represent three different slip systems, each with a Burgers vector $\frac{a}{2}[\bar{1}11]$. The slip planes are $(110)$, $(01\bar{1})$, and $(101)$. Allowing glide on three different slip planes makes cross slip possible and gives the distinct screw-edge differences seen in BCC materials \cite{feyPhasefieldDislocationModeling2022}. A BCC primitive cell is used to define the lattice grid points with primitive vectors $\mathbf{p}_1 = \frac{b}{\sqrt{3}}[11\bar{1}]$, $\mathbf{p}_2 = \frac{b}{\sqrt{3}}[\bar{1}11]$, and $\mathbf{p}_3 = \frac{b}{\sqrt{3}}[1\bar{1}1]$. In each simulation, the first order parameter is set to 0 or 1 depending on the initial dislocation configuration to create a dislocation on the $(110)$ plane. All other order parameters are initially zero. For the screw dislocation dipole simulations, a 128$b\times 362b\times 136b$ simulation cell is used, and the dislocations are initially 362$b$ long and separated by 32$b$. For the edge dislocation dipole simulations, a 128$b\times 128b \times 384b$ simulation cell is used, and the dislocations are also initially 362$b$ long and separated by 32$b$. In the dislocation loop simulations, a 128$b\times 128b \times 128b$ cell is used and the initial loop radius is 16$b$.

\section*{Acknowledgements}
LF acknowledges support from the Department of Energy National Nuclear Security Administration Stewardship Science Graduate Fellowship, which is provided under cooperative agreement number DE-NA0003960. SX and IJB gratefully acknowledge support from the Office of Naval Research under contract ONR BRC Grant N00014-21-1-2536. Use was made of computational facilities purchased with funds from the National Science Foundation (CNS-1725797) and administered by the Center for Scientific Computing (CSC). The CSC is supported by the California NanoSystems Institute and the Materials Research Science and Engineering Center (MRSEC; NSF DMR 1720256) at UC Santa Barbara.
HZ, XGL, CC, and SPO acknowledge support from the Office of Naval Research under Grant number N00014-18-1-2392 and computational resources provided by the University of California, San Diego, and the Extreme Science and Engineering Discovery Environment (XSEDE) supported by the National Science Foundation under grant no. ACI-1548562.
LQ acknowledges support from the National Science Foundation (NSF) under award DMR-1847837 and computational resources provided by Extreme Science and Engineering Discovery Environment (XSEDE) Stampede2 at the TACC through allocation TG-DMR190035.

\section*{Author Contributions}
HZ and XGL generated the DFT training data. HZ trained the ML potential and performed the atomistic calculations. LF designed and performed the PFDD calculations. YJH, LQ, and SX contributed the DFT training data of stacking faults for ternary and quaternary systems. LF and HZ wrote the manuscript with input from all authors. CC, SX, IJB, and SPO supervised the project. IJB and SPO designed the project.

\section*{Competing Information}
The authors declare no competing interests.

\section*{Additional Information}
The trained quaternary potential for MoTaNbTi is available in the Github repository of the open-source Materials Machine Learning (maml) package at
\url{https://github.com/materialsvirtuallab/maml/tree/master/mvl_models/pes}.

The training and test data are available at figshare with a DOI of 10.6084/m9.figshare.19247931. 

The PFDD code used in this study and the associated documentation is available under an open-source license at \url{https://github.com/lanl/Phase-Field-Dislocation-Dynamics-PFDD}.
\newpage

\bibliographystyle{unsrt}
\bibliography{refs}

\clearpage

\listoffigures
\end{document}


\title{Supplementary Information\\Multi-scale Investigation of Chemical Short Range Order and Dislocation Glide in the  MoNbTi and TaNbTi Refractory Multi-Principal Element Alloys}

\author[UCSD]{Hui Zheng \fnref{cor2}}
\author[UCSB]{Lauren T.W.\ Fey \fnref{cor2}}
\author[UCSD]{Xiang-Guo Li \fnref{cor2}}
\fntext[cor2]{These authors contributed equally}

\author[UM,Drexel]{Yong-Jie Hu}
\author[UM]{Liang Qi}
\author[UCSD]{Chi Chen}

\author[UCSBME]{Shuozhi Xu}
\author[UCSB,UCSBME]{Irene J.\ Beyerlein \corref{cor1}}
\ead{beyerlein@ucsb.edu}

\author[UCSD]{Shyue Ping Ong \corref{cor1}}
\ead{ongsp@eng.ucsd.edu}
\cortext[cor1]{Corresponding author}

\address[UCSD]{Department of NanoEngineering, University of California San Diego, La Jolla, CA 92093-0448, USA}
\address[UCSB]{Materials Department, University of California, Santa Barbara, CA 93106-5050, USA}
\address[UCSBME]{Department of Mechanical Engineering, University of California, Santa Barbara, CA 93106-5070, USA}
\address[UM]{Department of Materials Science and Engineering, University of Michigan, Ann Arbor, MI 48109, USA}
\address[Drexel]{Department of Materials Science and Engineering, Drexel University, Philadelphia, PA 19104, USA}
\date{}

\maketitle

\begin{figure}[H]
\centering\includegraphics[width=0.85\linewidth]{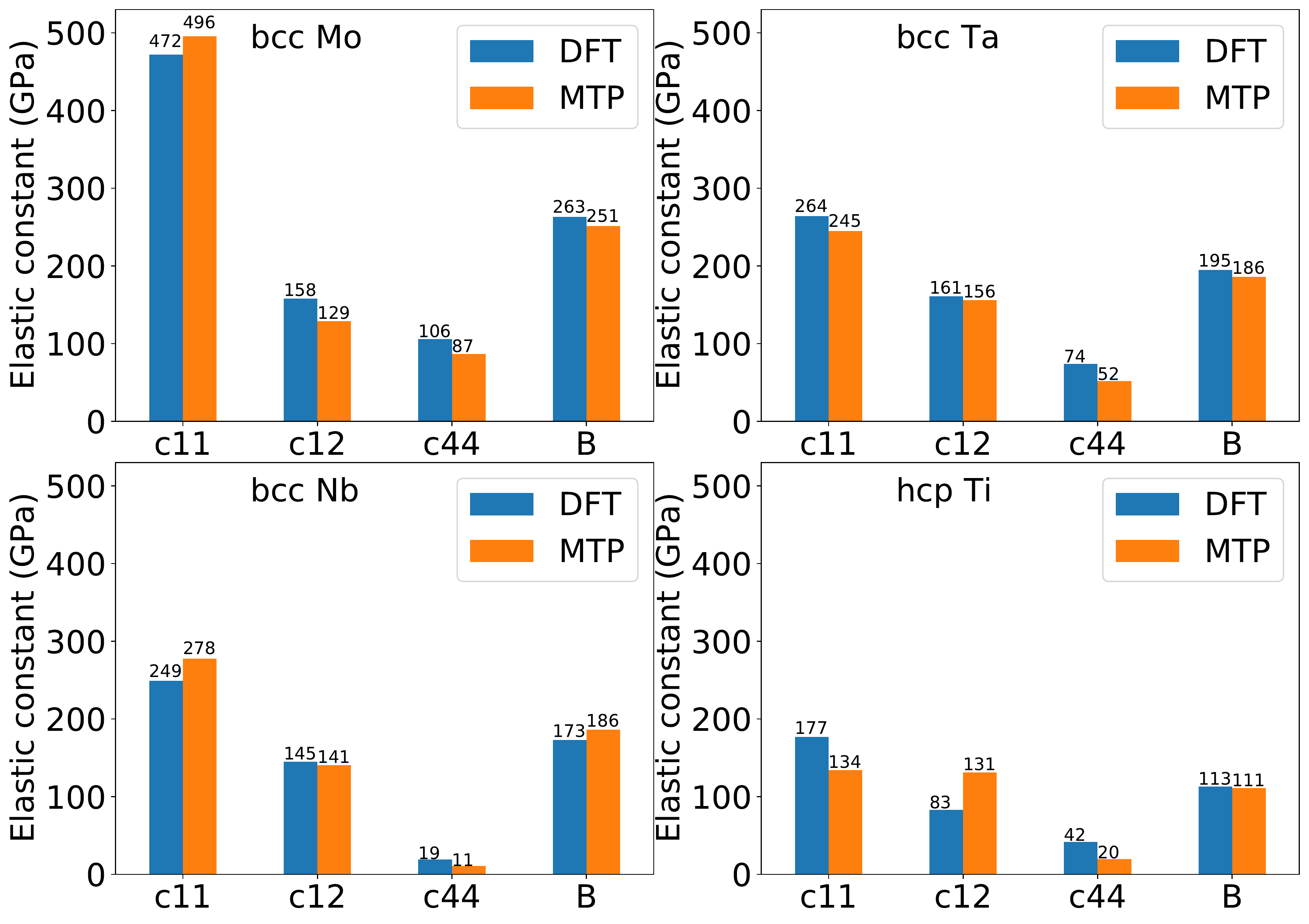}
\caption{\label{SI:elastic} Comparison of elastic constants of elemental systems calculated using density functional theory (DFT) and the trained moment tensor potential (MTP) for the Mo-Ta-Nb-Ti system. The values computed by MTP reproduce the DFT values for all elements.}
\end{figure}

\begin{figure}[H]
\centering\includegraphics[width=0.9\linewidth]{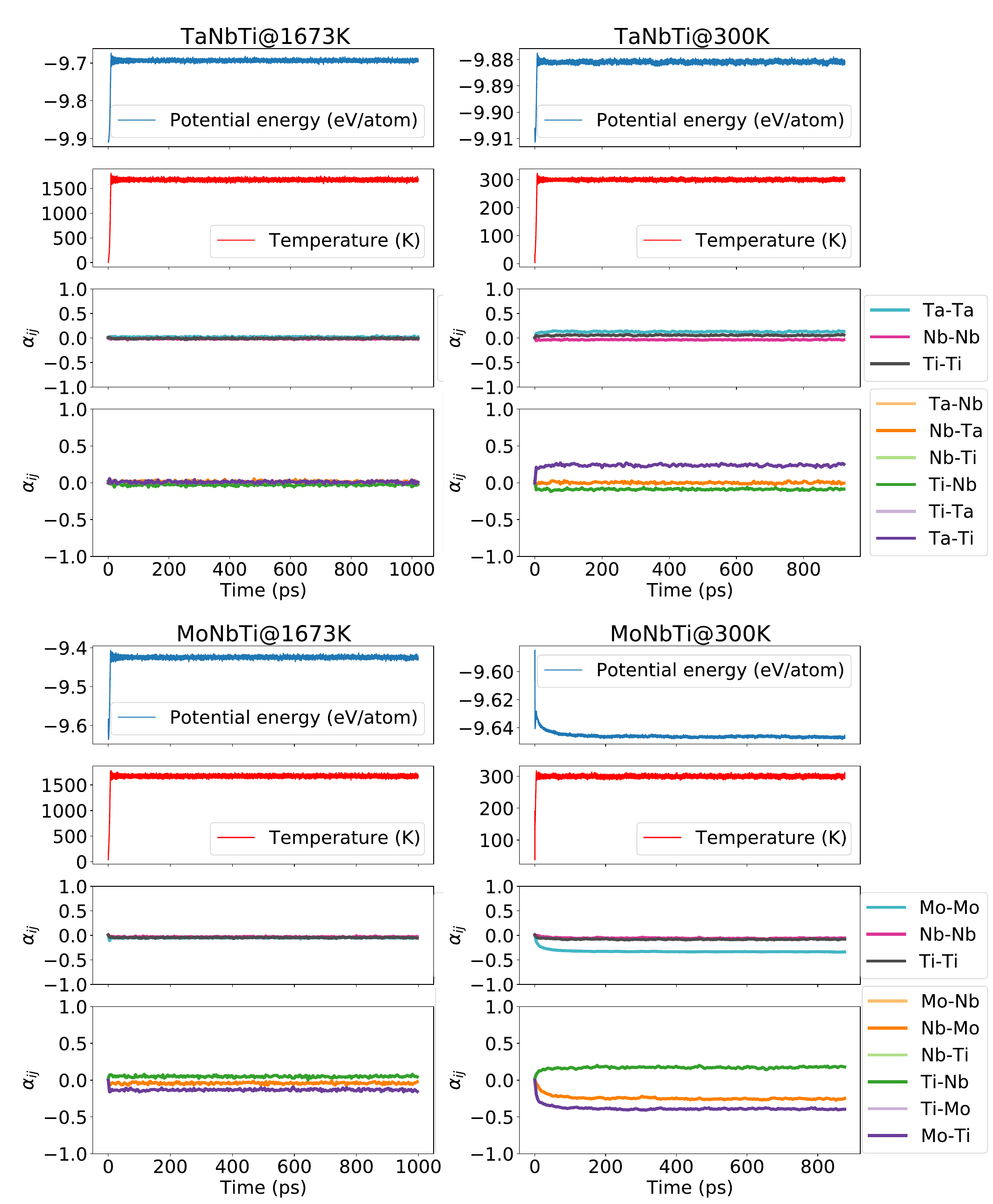}
\caption{\label{SI:SRO_vs_step_equi} The evolutions of potential energy, temperature, and SROs ($\alpha_{ij}$) of equimolar TaNbTi and MoNbTa from 300K and 1673K. With the cell size of 48 \AA $\times$ 46 \AA $\times$ 45 \AA\  of 5760 atoms, the energy and SRO converge after 200 ps.}
\end{figure}

\begin{figure}[H]
\centering\includegraphics[width=0.9\linewidth]{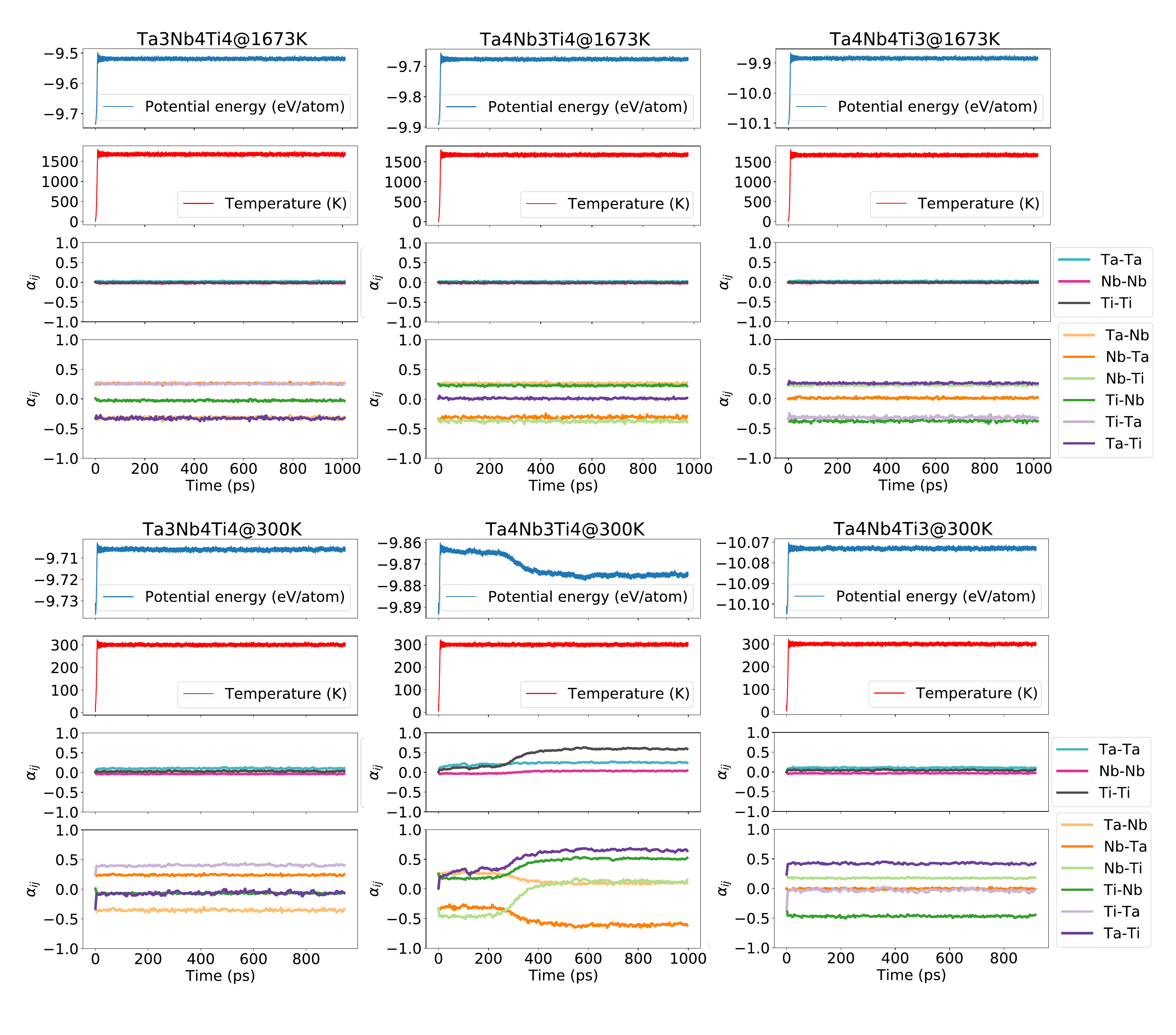}
\caption{\label{SI:SRO_vs_step_344_TaNbTi} The evolutions of potential energy, temperature, and SROs ($\alpha_{ij}$) from 300K and 1673K for Ta-Nb-Ti systems with a ratio of 3:4:4. With the cell size of 48 \AA $\times$ 46 \AA $\times$ 45 \AA\ of 5760 atoms, the energy and SRO converge after 200 ps for most cases except for \ce{Ta4Nb3Ti4}, for which energy and SRO converged after 600 ps. There is a significant bonding preference for the Ti-Ti and Nb-Ta pairs in \ce{Ta4Nb3Ti4}. The energy drops due to the formation of SRO is within 20 meV/atom in \ce{Ta4Nb3Ti4}. }
\end{figure}

\begin{figure}[H]
\centering\includegraphics[width=0.9\linewidth]{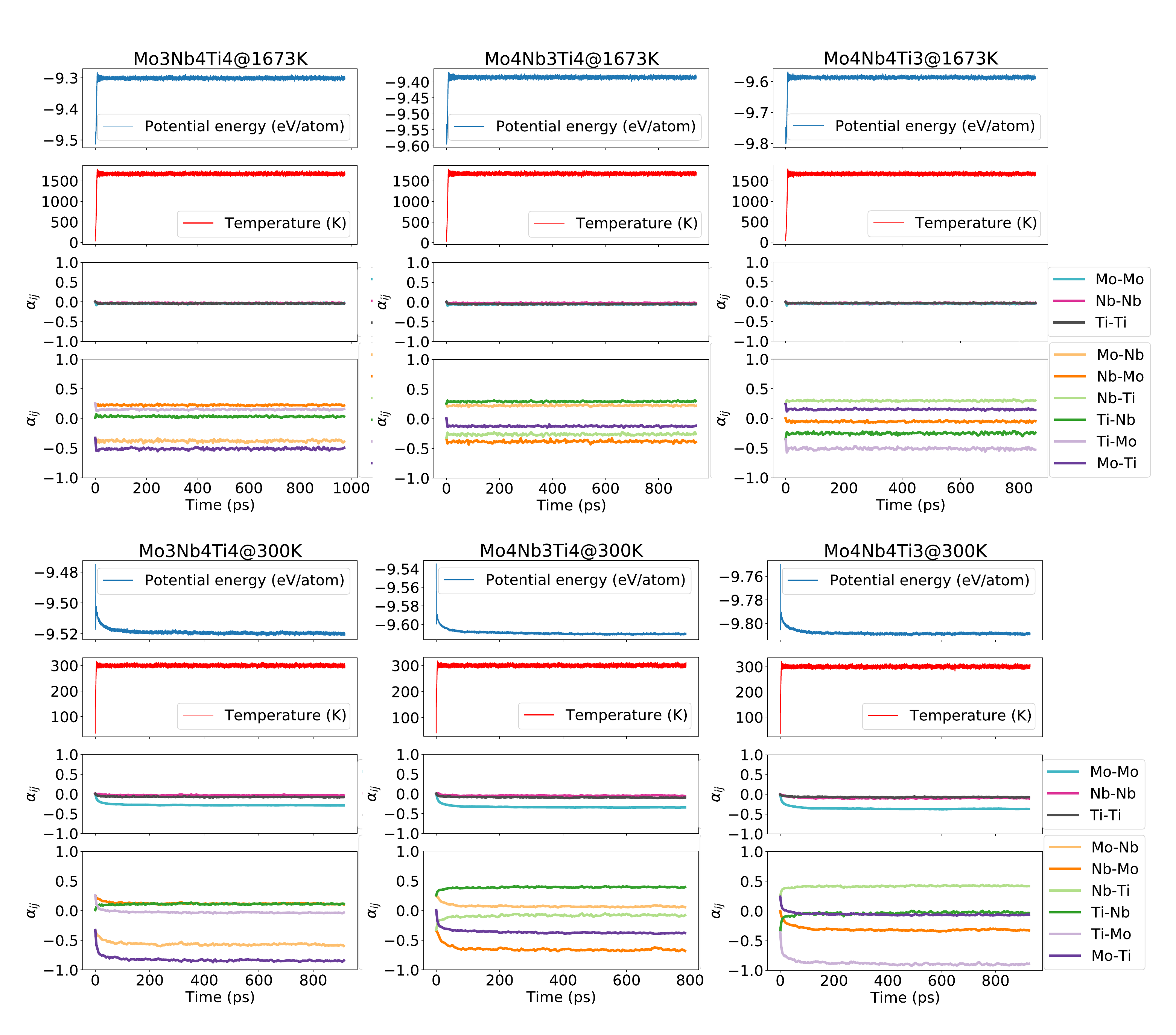}
\caption{\label{SI:SRO_vs_step_344_MoNbTi} The evolutions of potential energy, temperature, and SROs ($\alpha_{ij}$) from 300K and 1673K for Mo-Nb-Ti systems with a ratio of 3:4:4. With the cell size of 48 \AA $\times$ 46 \AA $\times$ 45 \AA\  of 5760 atoms, the energy and SRO converge after 200 ps for all cases.}
\end{figure}

\begin{figure}[H]
\centering\includegraphics[width=0.9\linewidth]{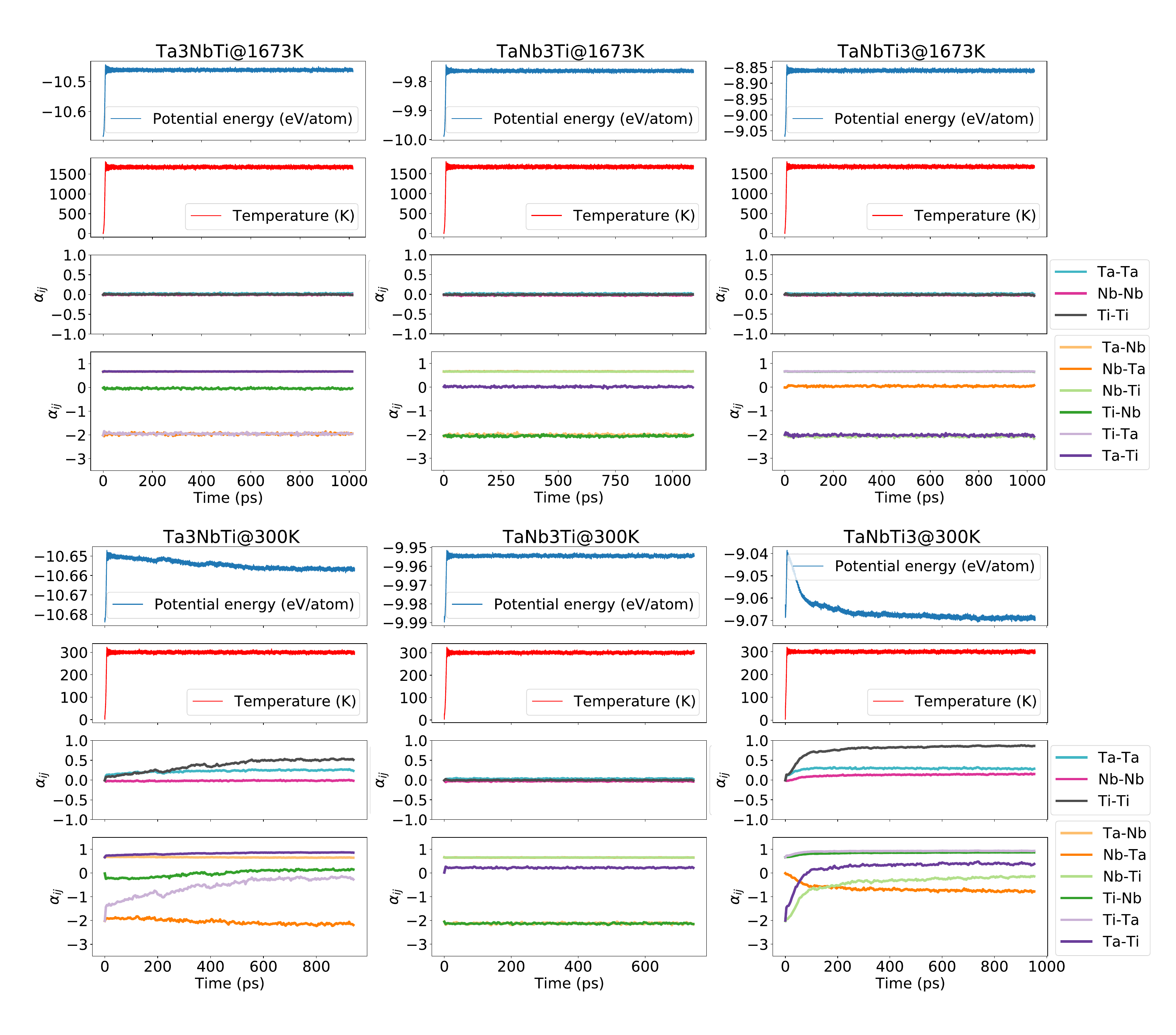}
\caption{\label{SI:SRO_vs_step_311_TaNbTi} The evolutions of potential energy, temperature, and SROs ($\alpha_{ij}$) from 300K and 1673K for Ta-Nb-Ti systems with a ratio of 3:1:1. With the cell size of 48 \AA $\times$ 46 \AA $\times$ 45 \AA\  of 5760 atoms, the energy and SRO converge after 200 ps for most cases except for \ce{Ta3NbTi} and \ce{TaNbTi3}, for these two, energy and SRO converged after 800 ps. There is a significant bonding preference for the Ti-Ti pair and Nb-Ta pairs in \ce{Ta3NbTi} and \ce{TaNbTi3}. The energy drops due to the formation of SRO is within 10 meV/atom in \ce{Ta3NbTi} and 10 meV/atom in \ce{TaNbTi3}.}
\end{figure}

\begin{figure}[H]
\centering\includegraphics[width=0.9\linewidth]{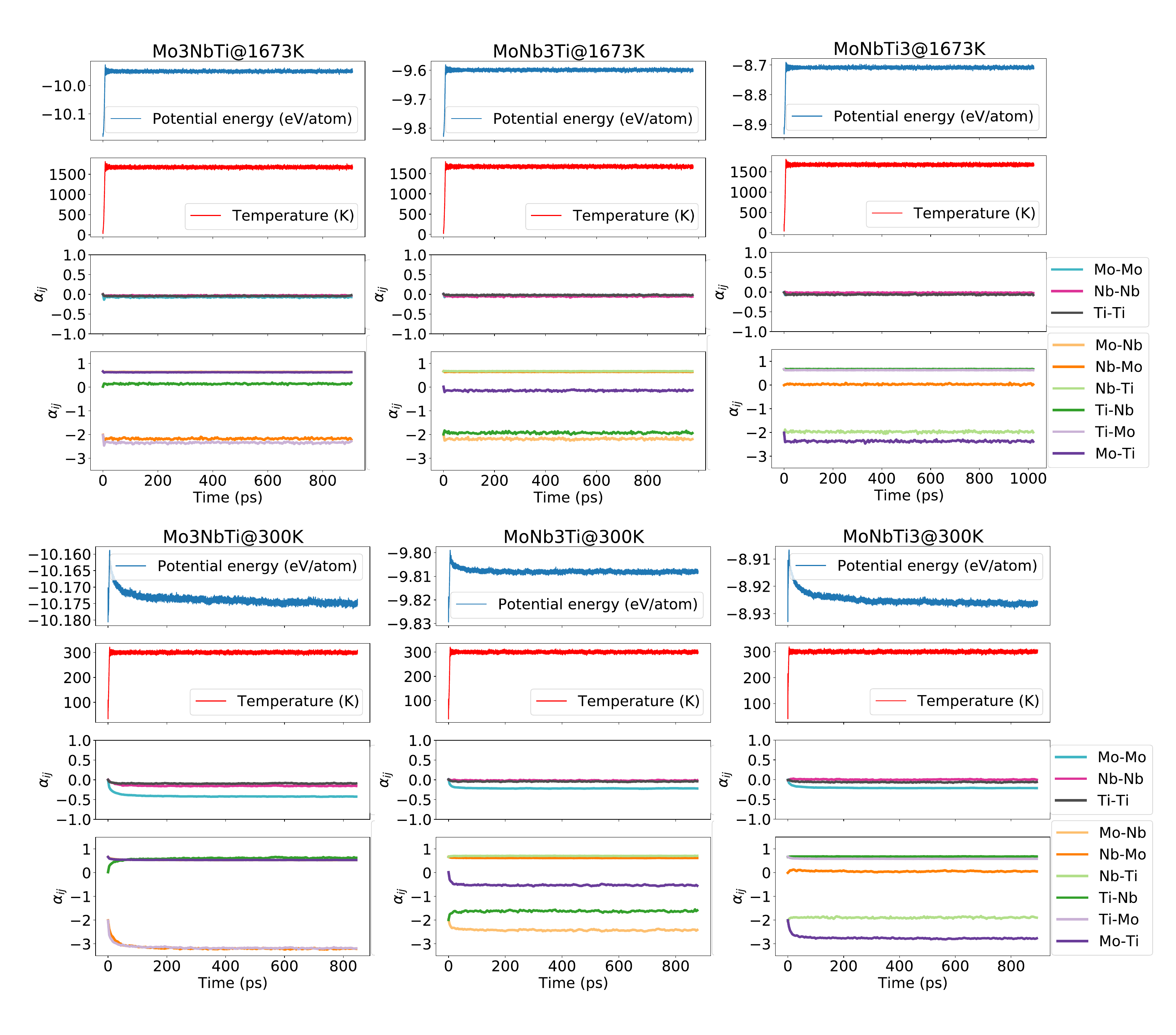}
\caption{\label{SI:SRO_vs_step_311_MoNbTi} The evolutions of potential energy, temperature, and SROs ($\alpha_{ij}$) from 300K and 1673K for Mo-Nb-Ti systems with a ratio of 3:1:1. With the cell size of 48 \AA $\times$ 46 \AA $\times$ 45 \AA\  of 5760 atoms, the energy and SRO converge after 200 ps for all cases. The non-equimolar composition significantly affects the preferred bonding pairs reflected in SROs. I.e., the atoms with a higher percentage prefer to surround and bond with those with a lower percentage. For example, in \ce{Mo3NbTi}, Nb-Mo pair and Ti-Mo are the most preferred bonding pairs. In \ce{MoNb3Ti}, Mo-Nb and Ti-Nb pairs are the preferred bonding pairs. Mo-Nb is slightly more preferred than the Ti-Nb pair. The energy drops due to the formation of preferred bonding are within 10 meV/atom for \ce{Mo3NbTi} and \ce{MoNb3Ti} and within 20 meV/atom for \ce{MoNbTi3}.}
\end{figure}

\begin{figure}
    \centering
    \includegraphics[width=0.9\linewidth]{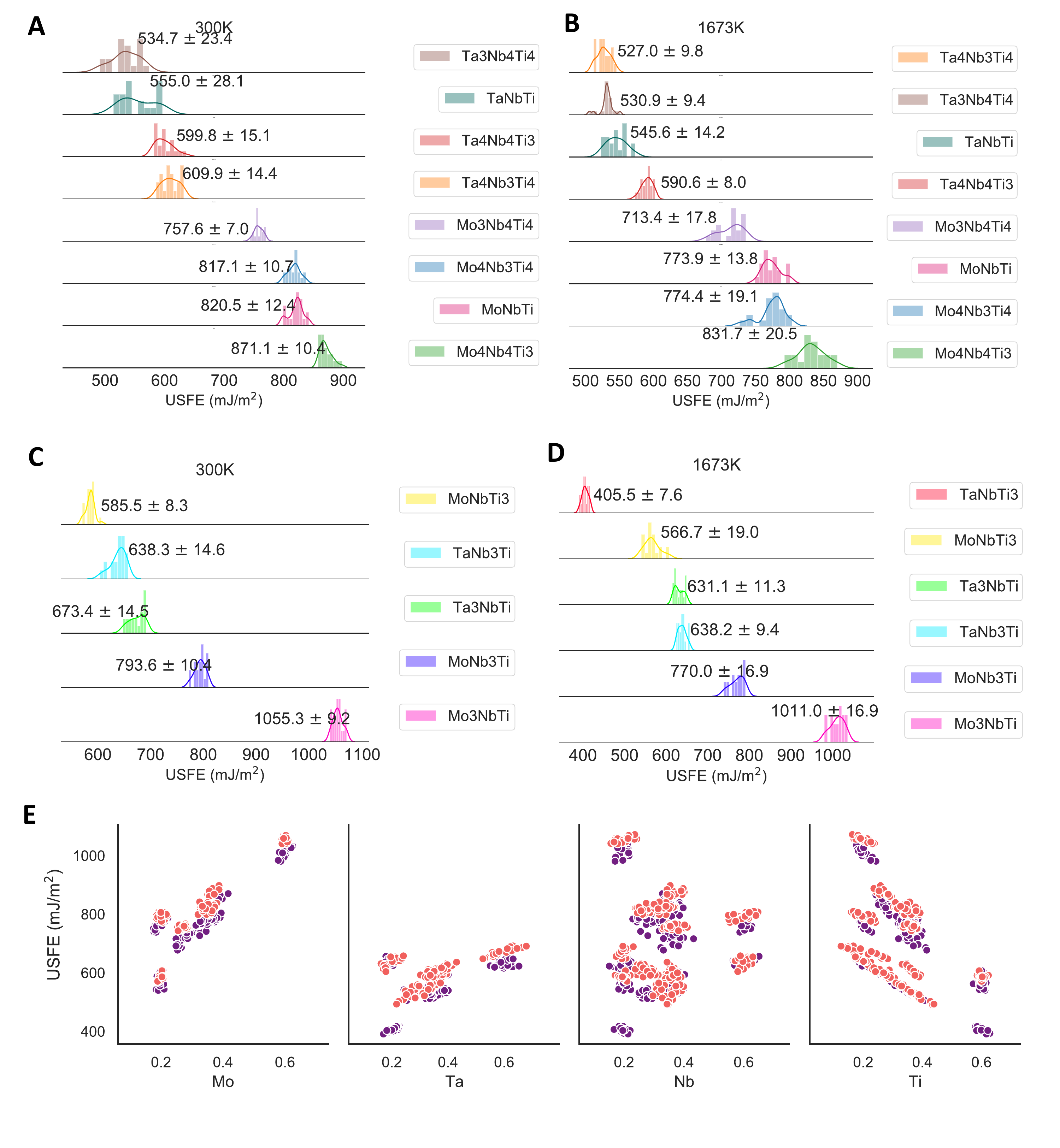}
    \caption{Stacking fault energy distribution of structures of different compositions annealed at 300 K and 1673 K.}
    \label{SI:SFE_dist}
\end{figure}

\begin{figure}
    \centering
    \includegraphics[width=0.85\linewidth]{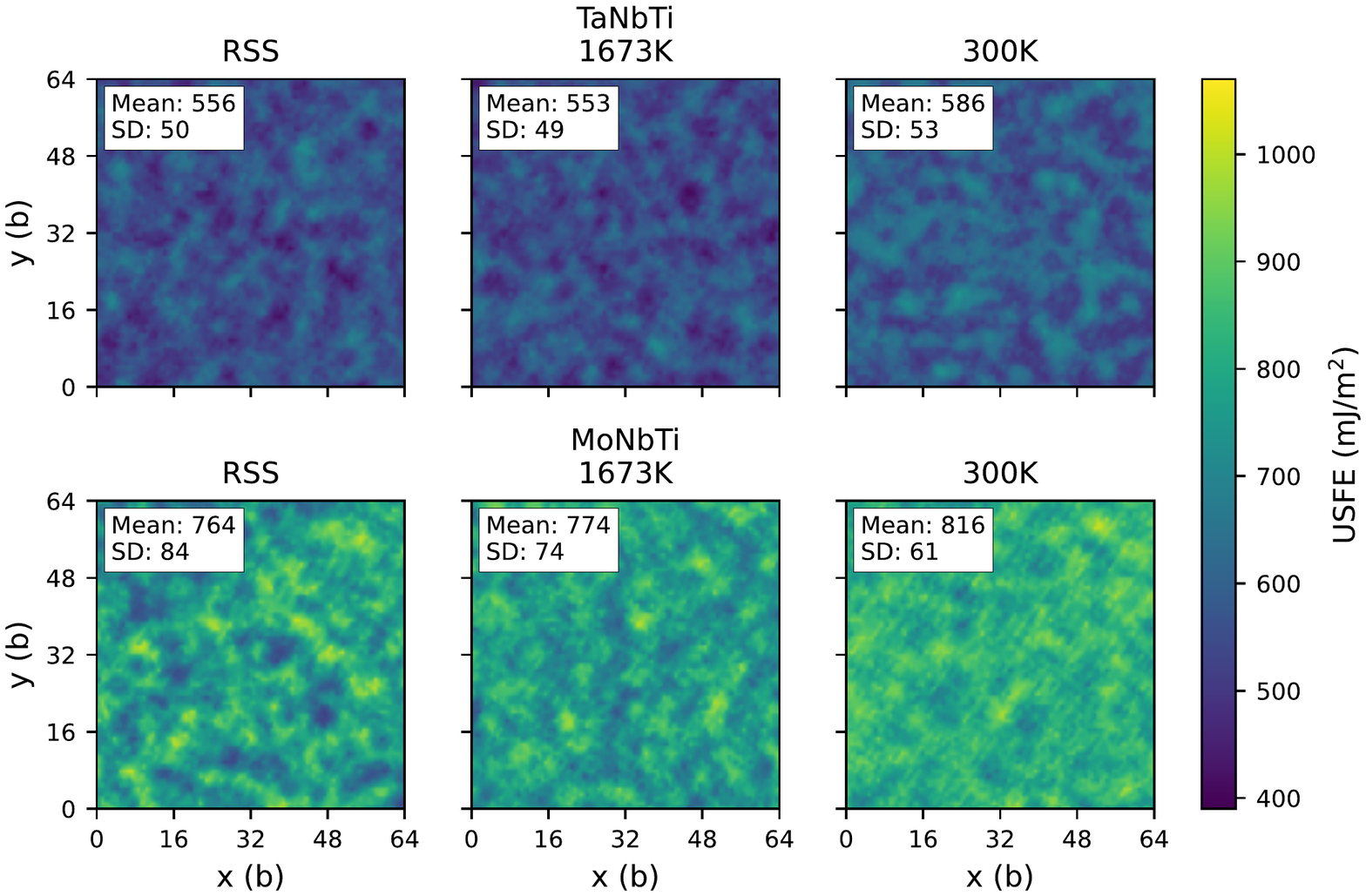}
    \caption{Example $(110)$ planes with colored by local unstable stacking fault energy (USFE) values for both alloys at different levels of short range order (SRO). The mean and standard deviation of the normally distributed USFEs are given in mJ/m$^2$.}
    \label{SI:usfe_surf}
\end{figure}

\begin{figure}
    \centering
    \includegraphics[width=0.85\linewidth]{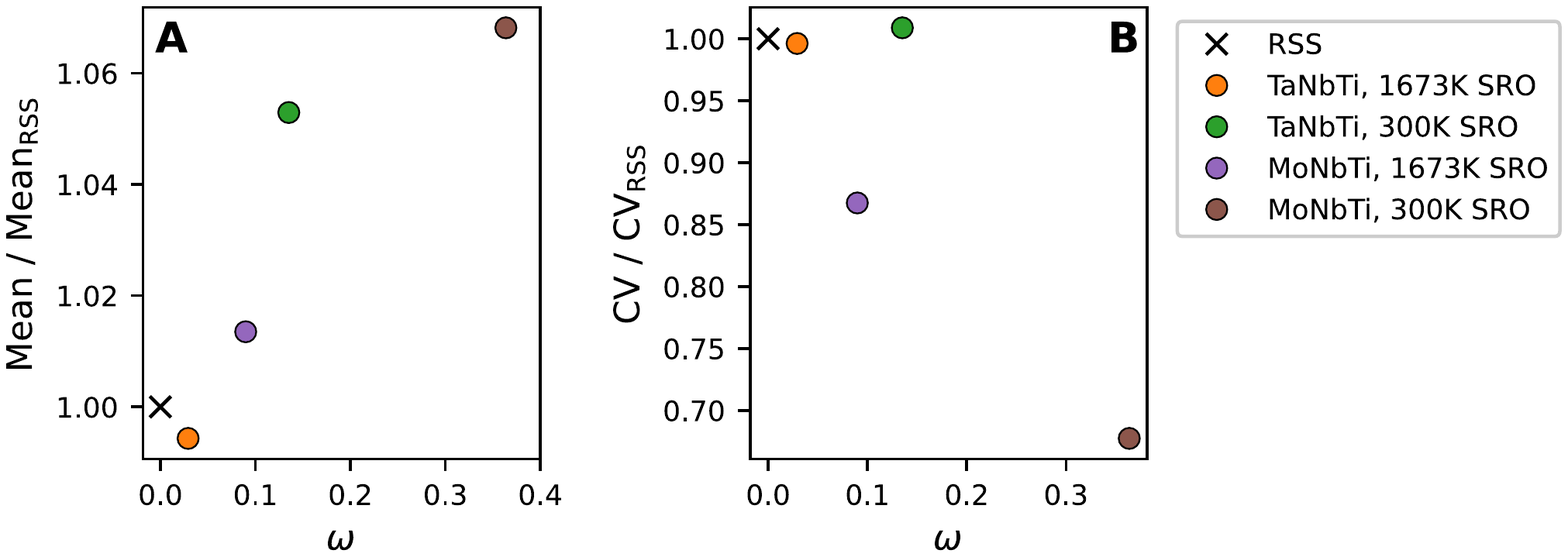}
    \caption{(A) The mean USFE for alloys with SRO normalized by the mean USFE for the random solid solution (RSS), plotted against the SRO figure of merit $\omega$. (B) The coefficient of variation in USFE for alloys with SRO normalized by the coefficient of variation in USFE for the no SRO case, plotted against the SRO figure of merit $\omega$. In both plots, the black $\times$ represents the RSS case. }
    \label{SI:usfe_fom}
\end{figure}